\documentclass[10pt]{revtex4}
\usepackage{amsmath}
\usepackage{bm}
\usepackage{bbm}
\usepackage{amssymb}
\usepackage{amsthm}
\usepackage{amsfonts}
\usepackage[latin1]{inputenc}
\usepackage{graphicx}
\usepackage{epstopdf}

\newtheorem{lem}{Lemma}
\newtheorem{prop}{Proposition}
\newtheorem{thm}{Theorem}
\newtheorem*{thm*}{Theorem}

\newtheorem{cor}{Corollary}

\numberwithin{equation}{section}

\def\TT{{\cal T}}

\def\RR{{\cal R}}
\def\RRR{{\mathbb R}}

\def\NNN{{\mathbb N}}
\def\LL{{\cal L}}

\def\FF{{\cal F}}
\def\CC{{\cal C}}

\def\HH{{\cal H}}

\def\II{{\cal I}}
\def\ZZZ{{\mathbb Z}}

\def\0{\emptyset}
\def\z{\zeta}
\def\L{\Lambda}

\def\b{\beta}
\def\m{\mu}
\def\d{\delta}

\def\r{\rho}

\def\s{\sigma}
\def\e{\varepsilon}

\def\G{\Gamma}
\def\p{\pi}

\def\o{\omega}

\def\O{\Omega}
\def\hbp{\hat {\bm p}}
\def\hp{\hat p}
\def\br{{\bm\r}}
\def\bof{{\bm f}}
\def\bz{{\bm z}}

\def\by{{\bm y}}

\def\bj{{\bm j}}
\def\bt{{\bm t}}
\def\bo{{\bm \o}}

\def\bp{{\bm p}}
\def\bze{{\bm\zeta}}
\def\bxi{{\bm\xi}}

\def\ol{\overline}

\def\dist{\operatorname{dist}}

\def\be{\begin{equation}}
\def\ee{\end{equation}}
\def\bea{\begin{eqnarray}}
\def\eea{\end{eqnarray}}

\def\nn{\nonumber}

\begin{document}

\title{
Evolution of correlation functions in the hard sphere dynamics
}

\author{Sergio Simonella}
\thanks{This work is a revised version of part of the author's PhD thesis \cite{S11}, 
written at the University of Rome ``La Sapienza'' under the direction of G. Gallavotti.}
\affiliation{INdAM--COFUND Marie Curie Fellow, \\ 
Zentrum Mathematik, TU M\"{u}nchen, \\
Boltzmannstrasse 3, 85748 Garching, Germany}

\begin{abstract}{The series expansion for the evolution of the correlation functions of a finite
system of hard spheres is derived from direct integration of the solution of the Liouville equation,
with minimal regularity assumptions on the density of the initial measure.
The usual BBGKY hierarchy of equations is then recovered. A graphical language based on the notion
of collision history originally introduced by Spohn is developed, as a useful tool for the description
of the expansion and of the elimination of degrees of freedom.
}\end{abstract}

\maketitle


\section{Introduction} \label{sec:int}

In his famous derivation of the Boltzmann equation \cite{Lanford}, O. E. Lanford makes use of a 
series expansion for the time--evolved correlation functions of a classical finite system of hard spheres in a box.
This expresses the $n-$point correlation function at time $t$ as a sum of integral terms involving all
the higher order correlation functions at time zero. The expansion is derived, though not rigorously,
from iteration of the BBGKY hierarchy of integro--differential equations, and is considered as a ``series
solution'' of its Cauchy problem. A rigorous validation of the hierarchy (formally deduced first by Cercignani in \cite{Cerc})
and of the series has been given years later by H. Spohn in an unpublished note \cite{Spohn}, and by R. Illner and 
M. Pulvirenti in \cite{IP} (see also the book \cite{CIP}), using different methods.

In both the previous papers an assumption on the initial measure is made to derive the BBGKY 
hierarchy, that is the continuity along trajectories of the hard spheres flow. 
However, there is no physical reason to expect such a regularity property to hold, and it is worthwhile 
to notice that the final series expansion makes perfectly sense without assuming it. In fact, Spohn observes 
at the end of his note, by a density argument, that the expansion can be extended to a more general class of measures 
having no continuity properties. On the other hand, the interpretation of the BBGKY hierarchy as a family
of partial differential equations is not at all easy, nor standard in any case, since it relies on the 
nontrivial properties of the operator $T_t$ of the hard sphere dynamics.
Hence, the series solution concept appears to be more appropriate for the description of the dynamics
in terms of probability distributions, and one wonders whether it is possible to derive it without going through
the usual hierarchy. The present paper is devoted to a derivation of the series expansion for the correlation 
functions, which is {\em not} based on the iteration of the BBGKY equations, 
and never requires continuity along trajectories. We rather construct a method of direct integration of the
solution of the Liouville equation, that allows to establish the validity of the expansion in a sense even 
{\em stronger} than those obtained in the existing literature: the result holds for all times in a fixed full measure 
invariant subset of the phase space, exactly as it happens for the existence of the dynamics of the underlying 
system of particles. The hierarchy of integro--differential equations is then recovered by resummation of the series, 
{\em without} additional assumptions on the initial measure, thus strengthening an analogous result in \cite{IP}.

Other rigorous discussions on the hard sphere dynamics and the associated BBGKY hierarchy are given in 
\cite{Uchiyama}, \cite{GeP}, \cite{PeG}, \cite{PGM} and \cite{GSRT13}.

Let us recall the derivation of Lanford and state our main result in an informal way. 
Consider the vector of correlation functions 
$\br = \{\r_n\}_{n \geq 1}$, where $\r_n$ is defined over the phase space of $n$ hard spheres of mass $m$
and diameter $a>0$ in a box $\L$. 
A point in this space is an $n-$tuple $(z_1,\cdots,z_n), z_j = (q_j,p_j)$, specifying position and momentum
of the $n$ particles. 
If $N$ is the total number of particles, we set $\r_n = 0$ for $n>N$. 
Then the BBGKY hierarchy for the evolution of $\r$ can be written
\be
\frac{\partial}{\partial t}\br(t) = H\br(t) + Q\br(t)\;, \label{eq:introBBGKY}
\ee
where
\be
\left(H \br\right)_n(z_1,\cdots,z_n,t) \equiv \Big\{H_n, \r_n\Big\}(z_1,\cdots,z_n,t) 
\ee
is the $n-$particles Liouville operator acting on $\r_n$ (including the effects of elastic collisions)
and the collision operator is defined by
\be
\left(Q\br \right)_n(z_1,\cdots,z_n,t) = a^2 \sum_{j=1}^n\int d\hat p\  d\o \ \o\cdot
\left(\frac{\hat p-p_j}{m}\right)\r_{n+1}(z_1,\cdots,z_n,q_j+a\o,\hat p,t)\;. \label{eq:introcoll}
\ee
Here $\hat p$ is integrated over all $\RRR^3$, and $\o$ runs over the unit sphere.

If $t \longrightarrow T_t(z_1,\cdots,z_n)$ is the flow of the dynamics, 
define the translation along trajectories of a vector of functions $\bof = \{f_n\}_{n \geq 1}$ as
\be
\left(S(t)\bof\right)_n(z_1,\cdots,z_n) = f_n(T_{-t}(z_1,\cdots,z_n))\;. \label{eq:introtrans}
\ee
Then, integration and iteration of Equation \eqref{eq:introBBGKY} leads to the formal solution
\be
\br(t) = S(t)\br(0) + \sum_{m=1}^{\infty}\int_0^t dt_1 \int_0^{t_1} dt_2\cdots \int_0^{t_{m-1}} dt_m
S(t-t_1) Q S(t_1-t_2)\cdots Q S(t_m) \br(0)\;. \label{eq:introresult}
\ee
In this paper {\em we analyze in detail the structure of Eq. \eqref{eq:introresult} and prove that it holds,
for all times in a full measure subset of the phase space, for any absolutely continuous measure with density
symmetric in the particle labels, and bounded by an equilibrium--like distribution.
}
The hierarchy \eqref{eq:introBBGKY} can be obtained then, in the mild sense of \cite{IP}, by taking the derivative.
No assumption of continuity is needed even for this last operation.
We also allow the total number of particles $N$ to be non fixed by the initial measure. 
The boundedness requirement is stronger than the necessary, and it is the 
same used by Lanford to control
the convergence of the series in the Boltzmann--Grad limit. Here it is made to control easily through all the steps
the integrals over momenta of the type \eqref{eq:introcoll}, \eqref{eq:introresult}.

The main interest of the discussion is the method of the proof. For $n=N$ Eq. \eqref{eq:introresult}
reduces to the evolution of the density function, that is the solution of the Liouville equation:
\be
\r_N(z_1,\cdots,z_N,t) = \r_N(T_{-t}(z_1,\cdots,z_N),0)\;. \label{eq:introliouv}
\ee
It is desirable that we can construct the series expansion for the $\r_n$ from {\em direct integration} of 
\eqref{eq:introliouv} over all the phase space of $N-n$ particles compatible with a fixed state $(z_1,\cdots,z_n)$. 
We show that in fact this can be done by eliminating the degrees of freedom particle by particle.
To achieve the integration of the single particle state, it is important to understand the structure
of the right hand side in \eqref{eq:introresult}. This has been widely studied since the work of Lanford \cite{Lanford},
see for instance \cite{King} or \cite{SpohnB}. It results that the integrand function in the generic term of the formula, 
depends on the states assumed by certain clusters of particles following a fictitious evolution: this is 
constructed from the state $(z_1,\cdots,z_n)$ at time $t$, by suitably adding more and more particles 
as the time flows backwards. Following \cite{SpohnB}, we shall call {\em collision history} such an evolution.

The collision histories can be represented graphically in terms of special binary tree graphs. Therefore,
a graphical picture of the series expansion \eqref{eq:introresult} is obtained. This representation is our basic
tool. In fact, it turns out that the integration of a particle state itself can be translated
in graphical language, through appropriate {\em operations over tree graphs}.
The graphical rules corresponding to the elimination of a particle state, clarify
how the various terms of the expansion for $\r_n$ emerge from those for $\r_{n+1}$, thus
considerably simplifying the presentation of the proof. The analytical operations corresponding to these
rules, are nothing but a suitable partitioning of the integration domain, and convenient representation 
(change of variables) of the subsets of the partition. 
Nevertheless, in order to establish the graphical rules, it is also essential to prove that some
classes of collision histories give a net null contribution to the integration of the particle state:
this is done again with the help of the tree graphs, by showing explicit one by one {\em cancellations}
among the collision histories of these classes. 

The paper is organised as follows. 
In Section \ref{sec:mod} we define the model, we introduce our notations and 
state our assumptions on the initial measure. In Section \ref{sec:coll} we introduce the concept of collision
history, as well as the graphical rules for its representation, and explain how to represent formula 
\eqref{eq:introresult} in terms of the tree graphs.
In Section \ref{sec:res} we present our main results, while in Section \ref{sec:proof} we discuss the 
proof of the main theorem, establishing the above mentioned graphical integration rules, and applying 
them to the generic inductive step.
In Section \ref{sec:conc} we present the conclusions.
A discussion on the hard sphere dynamics is deferred to the Appendix.


\section{The hard sphere system} \label{sec:mod}
In this section we set model and notations, which we inherit essentially from \cite{Spohn}, 
and state some preliminary result on the hard sphere dynamics (Section \ref{sec:mandn}). 
In Section \ref{subsec:ass} we introduce the class of measures on which we will work.

\subsection{Model and notations} \label{sec:mandn}

Let us consider a system of $N$ hard spheres of unit mass and of diameter $a>0$ moving in a box $\L\subset\RRR^3$.
$\L$ is bounded open and has a piecewise smooth elastically reflecting boundary $\partial\L$. We will
denote $z_i=(q_i,p_i)\in\L\times\RRR^3$ the configuration of the $i$--th particle, $i=1,\cdots,N$.
For groups of particles we will use the short notations $\bz_n = z_1,\cdots,z_n,$ $\bz_{n,j} = z_{n+1},\cdots, z_{n+j}$.
When there is no risk of confusion, we will simply call ``particle $i$'' a particle whose configuration is labelled by an index $i$.

We introduce the $n$ particle phase space, $n=1,\cdots,N$,
\be
\G_n = \Big\{\bz_n \in (\L\times\RRR^3)^n\ \Big|\ |q_i-q|\geq a/2 
\mbox{ for every }q\in\partial\L\ \mbox{and\ }|q_i-q_j|\geq a,\ i\neq j\Big\}\;.
\ee
A state of the system is given by a point in the full phase space $\G_N$. 

The equations of motion for the $n$ particle system are defined as follows.
Between collisions each particle moves on a straight line maintaining unchanged its velocity.
In a collision of two hard spheres at positions $q_i, q_j$ with $\o = (q_i-q_j)/|q_i-q_j|=(q_i-q_j)/a\in S^2$ and
with incoming momenta $p'_i, p'_j$ (that means $(p'_i-p'_j)\cdot\o <0$), we have instantaneous
transformation to the outgoing momenta $p_i, p_j$ (with $(p_i-p_j)\cdot\o >0$) given by
\bea
&&p_i = p'_i - \o[\o\cdot(p'_i-p'_j)]\;,\nn\\
&&p_j = p'_j + \o[\o\cdot(p'_i-p'_j)]\;.\label{eq:collpp}
\eea
%
Finally, in a collision of a particle with momentum $p'_i$ with the wall $\partial\L$ at a point
$q$ which is regular (there is only one point of contact between the wall and the sphere, and the 
normal to the surface at that point is well defined), we have instantaneous transformation to the 
reflected outgoing momentum $p_i$ given by 
\bea
p_i = p'_i - 2n(q)(n(q)\cdot p'_i)\;,\label{eq:collpw}
\eea
where $n(q)$ is the inner unit vector normal to $\partial\L$ in $q.$ It is easy to see that the collision 
transformations \eqref{eq:collpp} and \eqref{eq:collpw} are invertible and preserve Lebesgue measure on 
$\RRR^3\times\RRR^3$ and $\RRR^3$ respectively.

The above prescription for the equations of motion does not cover all possible situations, e.g. triple collisions
and collisions with corner points of the walls are excluded. Nevertheless, we have the following basic result:
\begin{prop} \label{prop:dyn} [Existence of the dynamics (I)]\ 
In $\G_n$ there is a subset $\G_n^*,$ whose complement is a Lebesgue null set, such 
that for any $\bz_n\in\G_n^*$ there is a unique mapping
\be
t \mapsto T_t^{(n)}\bz_n \in\G_n^*\ \ \ \ \ \ \ \ \ \ t \in\RRR\label{eq:nflow}
\ee
which is a solution of the equations of motion having $T_0^{(n)}\bz_n = \bz_n.$ Moreover, 
the shifts along trajectories $\bz_n\mapsto T_t^{(n)}\bz_n$ define a one--parameter group of 
Borel maps on $\G_n$ which leave Lebesgue measure invariant.
\end{prop}
\noindent This has been stated and proved by Alexander in \cite{Alexander}, pages 18--29,
and it holds under few simple regularity assumptions on $\partial\L$ (see pages 13--14 of \cite{Alexander} for the 
details on $\partial\L$). We shall make the same assumptions in the present paper.
The set $\G_n^*$ is shown to be a countable intersection of open sets with full measure.
The operator \eqref{eq:nflow} is called the flow of the $n$ particle dynamics. Another analysis of the hard
sphere dynamics may be found in \cite{MPPP}, \cite{CIP}.

Observe that (unlike in \cite{Alexander}) we do not identify ingoing and outgoing momenta of a collision, 
but we regard them as corresponding to distinct points in phase space, so that the flow $T^{(n)}_t$ is only 
piecewise continuous in $t$. When necessary, we distinguish the limit from the future $(+)$ and the limit 
from the past $(-)$ writing
\be
T^{(n)}_{t\pm}\bz_n = \lim_{\e\rightarrow 0^+} T^{(n)}_{t\pm\e}\bz_n\;.
\ee
For instance, in the statement of Proposition \ref{prop:dyn}, when $\bz_n \in \G_n^* \cap
\partial\G_n,$ it is understood that either $T_{0+}^{(n)}\bz_n = \bz_n$ or $T_{0-}^{(n)}\bz_n = \bz_n.$
From now on, to be more definite we fix the (irrelevant) convention
\be
T_t^{(n)}\bz_n = T_{t+}^{(n)}\bz_n\;. \label{eq:convfl}
\ee

The complement of $\G_n^*$ in $\G_n$ can be identified with the subset of points of $\G_n$ that evolved in time run into either:
\begin{itemize} 
\item a ``multiple'' collision, that is (i) simultaneous contact of more than two hard spheres, (ii) simultaneous contact
of two hard spheres with each other and at the same time with $\partial\L$ or (iii) simultaneous contact of one hard sphere
with two different points of $\partial\L$;
\item a grazing collision with the wall ($n(q)\cdot p'_i=0$) or a grazing two--body collision ($(p'_i-p'_j)\cdot\o = 0$);
\item a collision of a particle with a singular point $q\in\partial\L$ where the normal vector $n(q)$ is not well defined;
\item infinitely many collisions in finite time.
\end{itemize}
The flow through such situations will not be specified. We shall refer to them as the ``singular configurations''
(some examples in which a particle undergoes infinitely many collisions in a finite time are given in Sec. II.C
of \cite{Alexander}). 

We list some more notations that will be useful along the whole paper. For $\bz_n\in\G_n,$ we set
\bea
&& \G_{k}(\bz_n)=\Big\{\by_{k} \in (\L\times\RRR^3)^k\ \Big|\ (\bz_n,\by_k)\in\G_{n+k}\Big\}\;, \nn\\
&& \O_i(\bz_n)=\Big\{\o\in S^2\ \Big|\ (\bz_n,q_i+a\o,p)\in\G_{n+1}\ \ \forall p\in\RRR^{3} \Big\}\;,
\ \ \ \ \ \ \ \ \ \ \ i=1,\cdots,n\;.
\eea

To conclude this section, we pursue a bit further the analysis on the dynamics of the system of particles.
The following result will be used to study the properties of correlation functions:
\begin{prop} \label{prop:dynbis} [Existence of the dynamics (II)]
\ In $\G_n$ there is a subset $\G_n^{\dagger},$ whose complement is a Lebesgue null set, such 
that $\G_n^{\dagger} \subseteq \G_n^*,$ $T^{(n)}_t\G_n^{\dagger}=\G_n^{\dagger}$ and,
for $k=1,\cdots,N-n,$
\be
\bz_n\in\G_n^{\dagger}\ \Rightarrow\ (\bz_n,\bz_{n,k})\in\G_{n+k}^* \mbox{\ for a.a. $\bz_{n,k}\in\G_k(\bz_n)$}\;.
\ee
\end{prop}
\noindent This statement is a consequence of Proposition \ref{prop:dyn}. Its proof is given in the Appendix. 
Notice that $\bz_n\in\G_n^{\dagger}$ implies also $\bz_{n+k}\in\G_{n+k}^{\dagger}$
for a.a. $\bz_{n,k}$. The set $\G_n^{\dagger}$ can be identified with
%
\be
\G_n^{\dagger}= \Big\{\bz_n \in\G_n^* \ \Big|\ (T^{(n)}_{s}\bz_n, \bz_{n,k}) \in \G_{n+k}^*
\ \forall s \mbox{ and a.a. } \bz_{n,k}\in\G_k(T^{(n)}_{s}\bz_n)\Big\}\;, \label{eq:Gndagmax}
\ee
which is also the maximal subset obeying the properties of Proposition \ref{prop:dynbis}.

\subsection{Measures over the phase space} \label{subsec:ass}

Since all the particles of the system are identical, we will work with
the space $\mathcal L_N$ of Borel measurable functions $f_N:\G_N\rightarrow\RRR$, 
symmetric in the particle labels
($f_N(\Pi (z_1,\dots,z_N)) = f_N(z_1,\dots,z_N)$ for any permutation $\Pi$). We also assume that the 
functions in $\mathcal L_N$ have a boundedness property on $\G_N$ of the type
\be
|f_N(\bz_N)| \leq A \prod_{j=1}^N h_{\b}(p_j)\;,\ \ \ \ \ \ \ \ \ \ \ \ \ \ \ 
h_{\b}(p) =\left( \frac{\b}{2\p}\right)^{\frac{3}{2}}e^{-\frac{\b}{2}p^2}\;,\label{eq:assbound}
\ee
for some $A,\b >0$. 

In Eq. \eqref{eq:assbound} we ignore the (possible) dependence on $N$ of the constant $A$, being the total number of particles always 
fixed throughout the paper. In particular, we allow $A$ to grow exponentially with $N$. It is worth to notice that this includes the 
states considered in the derivation of the Boltzmann equation~\cite{Lanford}.


Suppose to have an initial measure $P$ on $\G_N$ with density $f_N^0 \in\mathcal L_N$
with respect to the Lebesgue measure $d\bz_N=dz_1\dots dz_N$, 
\bea
P(d\bz_N)=f_N^0(\bz_N)d\bz_N\;.
\eea
Then, because the flow $T^{(N)}_t$ preserves the Lebesgue measure,
the evolved measure at time $t$ has a density $f_N(t)$ given by
\be
f_N(\bz_N,t) = f_N^0(T^{(N)}_{-t}\bz_N) \label{eq:liouville}
\ee
almost everywhere in $\G_N$, which is the Liouville equation in mild form.
Points of $\G_N \setminus \G_N^*$ are removed from \eqref{eq:liouville}. 
Notice that estimate \eqref{eq:assbound} is preserved in the time evolution by conservation of energy.
In particular, $f_N(t)\in \mathcal L_N$.
Of course since the flow $T^{(N)}_t$ is only well defined almost surely, even densities that are smooth 
at time zero will only be $\mathcal L_N-$functions at time $t$.
Observe that, by our convention \eqref{eq:convfl}, $f_N$ takes the same value in incoming and outgoing states of collision
(a property not to be confused with the continuity along trajectories; see Eq. \eqref{eq:conttraj} below).

We define the correlation functions $\r_n(t)\in\mathcal L_n, n=1,2,\dots$ by
\bea
&&\r_n(\bz_n,t)
= N \dots (N-n+1) \int_{\G_{N-n}(\bz_n)}dz_{n+1}\dots dz_N f_N(\bz_N,t)\;,\ \ \ \ \ n\leq N\;,\nn\\
&&\r_n = 0\;,\ \ \ \ \ \ \ \ \ \ n>N\;,\nn\\
&&\r_n^0(\bz_n)\equiv\r_n(\bz_n,0)\;.  \label{eq:defcf}
\eea
$P$ can be, in general, any signed measure with density in ${\cal L}_N.$
In the case $P$ is a probability measure, the quantity
\be
\frac{1}{N\cdots (N-n+1)}\int_{\cal W}dz_1\cdots dz_n \r_n(z_1,\cdots,z_n,t)
\ee
is the probability of finding particles $1,2,\cdots,n$ at time $t$ in the Borel set ${\cal W}\in\G_n$.

\section{Collision histories} \label{sec:coll}

In this section we analyze the structure of the expansion on the right hand side of \eqref{eq:introresult}
(Section \ref{subsec:fict}). This is given in general by a large variety of terms. 
In order to have a clear picture of the many terms of the expansion and of the configurations of particles involved in them, 
we shall establish rules for their graphical representation (Section \ref{subsec:family}).

From now on, without loss of generality and to avoid overweight of notation, the time $t$ will be always supposed to be positive. 


\subsection{The structure of formula \eqref{eq:introresult}: an integral over fictitious evolutions of particles} 
\label{subsec:fict}

Let us look carefully at the explicit expression of the right hand side in Eq. \eqref{eq:introresult}. We compute it, say, in $\bz_n$,
taking into account the definitions \eqref{eq:introcoll} and \eqref{eq:introtrans}. We see that, in the generic term, the integrand 
function contains one time--zero correlation function. This is evaluated in a configuration of particles which can be found by 
flowing backwards in time the configuration $\bz_n$, and suitably adding new particles at the times $t_1, t_2$ etcetera. 
The new particles appear in a collision configuration with one of the pre--existent particles. This describes a special evolution 
that will be called ``collision history'', a name first used by Spohn in \cite{Spohn}.

We want to stress since the beginning that the collision history is {\em not} a real trajectory of the particle system, and the associated 
collisions are not a sequence of real collisions. The correspondence between collision histories and sequences of real collisions 
is only very indirect (\cite{Spohn}).

We begin by explaining how to construct a {\em collision history}. The ingredients are the collection of variables (in parentheses 
we specify what will be their interpretation):
\begin{itemize} \label{list:ch}
\item $n\in\{1,2,3,\cdots\}$\ \ \ \ \ \ \ \ \ \ \ \ \ \ \ \ \ \ \ \ \ \ \ \ \ \ \ \ \ \ \ \ \ \ \ \ \ \ \ \ \ \ \ \ \ \ \ \ 
\ \ \ \ \ \ \ \ \ \ \ \ \ \ \ \ \ \ \ \ \ \ \ \ \ (starting number of particles),
\item $m\in\{0,1,2, \cdots \}$\ \ \ \ \ \ \ \ \ \ \ \ \ \ \ \ \ \ \ \ \ \ \ \ \ \ \ \ \ \ \ \ \ \ \ \ \ \ \ \ \ \ \ \ \ \ \ \ \ \ \ \ \ \ \ \ \ \ \ \ \ \ \ \ 
\ \ \ \ \ \ \ \ \ \ \ (number of added particles),
\item $\bz_n\in\G_n^*$\ \ \ \ \ \ \ \ \ \ \ \ \ \ \ \ \ \ \ \ \ \ \ \ \ \ \ \ \ \ \ \ \ \ \ \ \ \ \ \ \ \ \ \ \ \ \ 
\ \ \ \ \ \ \ \ \ \ \ \ \ \ \ \ \ \ \ \ \ \ \ \ \ \ \ \ \ \ \ \ \ \ \ \ \ \ \ \ \ \ \ \ \ \ \ \ \ \ \ (starting configuration),
\item $t>0$\ \ \ \ \ \ \ \ \ \ \ \ \ \ \ \ \ \ \ \ \ \ \ \ \ \ \ \ \ \ \ \ \ \ \ \ \ \ \ \ \ \ \ \ \ \ \ \ \ \ \ \ \ \ \ \ \ \ \ \ \ \ \ \ 
\ \ \ \ \ \ \ \ \ \ \ \ \ \ \ \ \ \ \ \ \ \ \ \ \ \ \ \ \ \ \ \ \ \ \ \ \ \ \ \ \ \ \ \ \ \ \ \ (total time span),
\item $\bt_m\in\RRR^m\ (m\geq 1)\ $ with $t\equiv t_0>t_1>\cdots>t_m>t_{m+1}\equiv 0$ 
\ \ \ \ \ \ \ \ \ \ \ \ \ \ \ \ \ \ \ \ \ \ \ \ \ \ \ \ \ \ \ \ \ \ \ (times of creation 

\ \ \ \ \ \ \ \ \ \ \ \ \ \ \ \ \ \ \ \ \ \ \ \ \ \ \ \ \ \ \ \ \ \ \ \ \ \ \ \ \ \ \ \ \ \ \ \ \ \ \ \ \ \ \ \ \ \ \ \ \ \ \ \ \ \ \ \ \ \ \ \ \ \ \ \ \ \ \ \ \ \ \ \ \ 
\ \ \ \ \ \ \ \ \ \ \ \ \ \ \ \ \ \ \ \ \ \ \ \ \ \ \ \ \ \ \ \ \ \ of added particles),
\item $\bj_m\in\NNN^m\ (m \geq 1)\ $ with $j_1\in I_{n},\cdots, j_m\in I_{n+m-1},$ where $I_{k} = \{1,\cdots,k\}$ 
\ \ \ \ \ \ \ \ \ \ \ \ \ \ \ \ \ (progenitors of 

\ \ \ \ \ \ \ \ \ \ \ \ \ \ \ \ \ \ \ \ \ \ \ \ \ \ \ \ \ \ \ \ \ \ \ \ \ \ \ \ \ \ \ \ \ \ \ \ \ \ \ \ \ \ \ \ \ \ \ \ \ \ \ \ \ \ \ \ \ \ \ \ \ \ \ \ \ \ \ \ \ \ \ \ \ 
\ \ \ \ \ \ \ \ \ \ \ \ \ \ \ \ \ \ \ \ \ \ \ \ \ \ \ \ \ \ \ \ \ \ \ \ \ \ \ \ \ added particles),
\item $\hbp_m\in\RRR^{3m}\ (m\geq 1)$\ \ \ \ \ \ \ \ \ \ \ \ \ \ \ \ \ \ \ \ \ \ \ \ \ \ \ \ \ \ \ \ \ \ 
\ \ \ \ \ \ \ \ \ \ \ \ \ \ \ \ \ \ \ \ \ \ \ \ \ \ \ \ \ \ \ \ \ \ \ \ \ \ \ \ \ \ \ \ (momenta of added particles 

\ \ \ \ \ \ \ \ \ \ \ \ \ \ \ \ \ \ \ \ \ \ \ \ \ \ \ \ \ \ \ \ \ \ \ \ \ \ \ \ \ \ \ \ \ \ \ \ \ \ \ \ \ \ \ \ \ \ \ \ \ \ \ \ \ \ \ \ \ \ \ \ \ \ \ \ \ \ \ \ \ \ \ \ \ \ \ \ \ \ \ \ \ \ \ \ 
at the time of their creation),
\item $\bo_m\in S^{2m}\ (m \geq 1)\ $, with a constraint defined below\ \ \ \ \ \ \ \ \ 
(relative position, in units of $a,$ of the added 

\ \ \ \ \ \ \ \ \ \ \ \ \ \ \ \ \ \ \ \ \ \ \ \ \ \ \ \ \ \ \ \ \ \ \ \ \ \ \ \ \ \ \ \ \ \ \ \ \ \ \ 
\ \ \ \ \ \ \ \ \ \ \ \ \ \ \ \ \ \ \ \ \ \ \ \ \ \ \ \ \ \ \ \ \ \ \ \ particles with respect to their progenitors).
\end{itemize}

\begin{figure}[htbp] 
   \centering
   \includegraphics[width=3in]{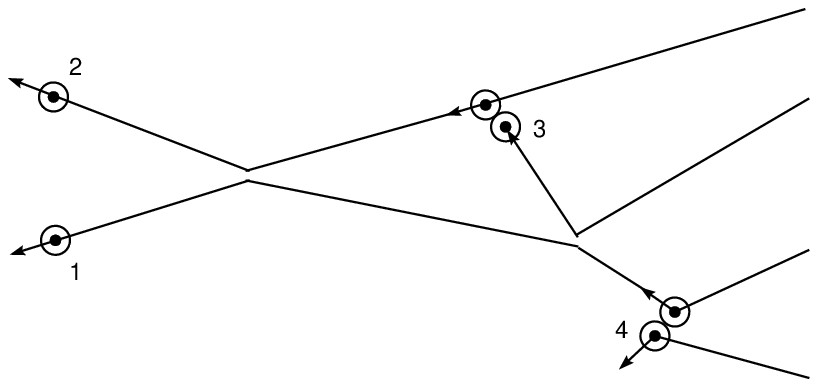} 
   \caption{Trajectory drawn by the particles in a collision history, in the case 
   $n=2, j_1=2, j_2=1$. Here particle $3$ is added in an incoming collision configuration ($\o_1\cdot (\hat p_1-\p_{j_1}(t_1))<0$), 
   while particle $4$ is added in an outgoing collision configuration ($\o_2\cdot (\hat p_2-\p_{j_2}(t_2))>0$).}
   \label{fig:traj}
\end{figure}

To any choice of the variables in the list we associate a backwards evolution. We indicate with the greek letter
\be
\z_i(s) = (\xi_i(s),\p_i(s)) \in \L\times\RRR^3 
\ee
the configuration of particle $i$ (position and momentum) at time $s$ in such evolution,
defined as follows. Take the starting configuration $\bz_n\in\G_n^*$, put $(\z_1(t),\cdots,\z_n(t)) = \bz_n$, 
and evolve it backwards in time as if there were no other 
particles in the space up to time $t_1.$ This defines the piecewise continuous trajectory $(\z_1(s),\cdots,\z_n(s))$ 
for $t_1 < s < t,$ that is $(\z_1(s),\cdots,\z_n(s)) = T^{(n)}_{-t+s}\bz_n.$ Set $(\z_1(t_1),\cdots,\z_n(t_1)) =  T^{(n)}_{-t+t_1}\bz_n.$
If $m=0,$ put $t_1=0:$ the construction is finished. Otherwise, at time $t_1$ stop your $n$ particle system and 
add particle $n+1$ in a state $\z_{n+1}(t_1)=(\xi_{j_1}(t_1)+a\o_1,\hp_{1}),$ with $\o_1\in\O_{j_1}(\bze_n(t_1))$ and such
that the dynamics of the obtained system of $n+1$ particles is well defined, i.e. $(\bze_n(t_1),\z_{n+1}(t_1))\in\G_{n+1}^*.$ 
Observe that, at fixed $\bz_n, t_1,$ we will have either an incoming or an outgoing 
collision between particles $j_1$ and $n+1$, depending on the chosen values of $\o_1, \hp_{1}$. 
Now, evolve backwards in time particles $1,\dots,n+1$
as if there were no other particles in the space up to time $t_2<t_1:$ this defines the piecewise continuous trajectory 
$(\z_1(s),\cdots,\z_{n+1}(s)) = T^{(n+1)}_{-t_1+s}\bze_{n+1}(t_1)$ for 
$t_2 < s < t_1.$ Notice that, soon after $t_1,$ particle $j_1$ in the backwards evolution
will deviate from its free motion if and only if $\o_1,\hp_{1}$ correspond to an outgoing collision.
Set $\bze_{n+1}(t_2) =  T^{(n+1)}_{-t_1+t_2}\bze_{n+1}(t_1).$
If $m=1$ ($t_2=0$) the construction is finished. Otherwise, at time $t_2$ stop the system and add particle $n+2$ 
as above with momentum $\hat p_{2}$ and position at distance $a\o_{2}$ from particle $j_2,$
with $\o_2\in\O_{j_2}(\bze_{n+1}(t_2))$ and the constraint that the obtained system of $n+2$ particles is in $\G_{n+2}^*.$
Later on evolve your $n+2$ particles backwards in time up to time $t_3<t_2$, and so on
up to the final step, which is the evolution of particles $1,\dots,n+m$ with the flow $T^{(n+m)}_{-t_{m}+s}, 0\leq s<t_{m}.$ 
We shall say in the future that particle $n+k$ is ``created'' by particle $j_k$, or that particle $j_k$ is its ``progenitor''.
An example is pictured in 
 \ref{fig:traj}.

We will use always greek alphabet for collision histories. We will call $\bze(s)$ the configuration of all the particles of 
the history at time $s.$ When no confusion arises, this symbol will have {\em no} subscript specifying the number of particles,
which is actually variable in time, so that 
\be
\bze(s) = (\bxi(s),\bp(s)) = (\xi_1(s),\cdots,\xi_{n+k}(s),\p_1(s),\cdots,\p_{n+k}(s))\ \ \ \ \ \ \ \ \ \ \ 
\mbox{for $s\in(t_{k+1},t_k]$}\;.
\ee
In particular, if $s$ coincides with a time $t_k$, then $\bze(s)$ 
is the configuration of the particles of the evolution {\em after} having added the new particle $n+k$ (but {\em before} the
related backwards collision, in the case the added particle is in outgoing configuration): 
\be
\bze(t_k) = (\bze_{n+k-1}(t_k),\xi_{j_k}(t_k)+a\o_k,\hp_k)\;.
\ee
%

Now we turn back to the description of the series expansion. A careful look of formula \eqref{eq:introresult}
leads to the following more explicit formal representation:
\be
\r_n(\bz_n,t) = \sum_{m=0}^{\infty}\sum_{\substack{j_1,\cdots,j_m\\ j_k\in I_{n+k-1}}}
\int_{\CC_{\bj_m}(\bz_n,t)} d\m(\bt_m,\bo_m,\hat\bp_m) \left(\prod_{k=1}^m 
B(\o_k; \hp_k-\p_{j_k}(t_k))\right)
\r_{n+m}^0(\bze(0))
\label{eq:seriesrev}
\ee
where
\bea
&& d\m(\bt_m,\bo_m,\hat\bp_m) \equiv d\m_m = dt_1\cdots dt_md\o_1\cdots d\o_m d\hp_1\cdots d\hp_m\nn\\
&&\ \ \ \ \ \ \ \ \ \ \ \ \ \ \ \ \ \ \ \ \ \ \ \ \ \ \ \ \ \ 
\mbox{(= volume element over $\RRR^m\times S^{2m}\times\RRR^{3m}$)}\;,\nn\\
&& B(\o_k; \hp_k-\p_{j_k}(t_k)) \equiv B_k = a^2\o_k \cdot \left(\hp_k-\p_{j_k}(t_k)\right)\;, \nn\\
&& \CC_{\bj_m}(\bz_n,t) = \Big\{(\bt_m,\bo_m,\hat\bp_m)\in\RRR^m\times S^{2m}\times\RRR^{3m}\ \Big|\ \nn\\
&& \ \ \ \ \ \ \ \ \ \ \ \ \ \ \ \ \ \ \ \ t>t_1>\cdots>t_m>0,\ \o_k\in\O_{j_k}(\bze_{n+k-1}(t_k))\Big\}\;.
\label{eq:notterm}
\eea
The volume element $d\m_m$ is the induced Lebesgue measure over $\RRR^m\times S^{2m}\times\RRR^{3m}$.
The sum over $m$ is extended to infinity by the convention in \eqref{eq:defcf}. The term $m=0$ must be interpreted 
as $\r_{n}^0(T^{(n)}_{-t}\bz_n)$.

In the following sections we will rigorously derive \eqref{eq:seriesrev} from Liouville equation,
starting with an initial density $f_N^0\in \mathcal L_N.$ In particular,
we will check its consistency, and this will require to prove that the collision histories involved in the integrand
are well defined  $d\m_m-$a.e. in the domain of integration. 

Observe that, since $\r_{n+m}^0$ satisfies an estimate as in \eqref{eq:assbound}, applying conservation of energy
at each creation of the collision history we find
\be
\r_{n+m}^0(\bze(0)) \leq A' \prod_{j=1}^n h_\b(p_j)\prod_{k=1}^m h_\b(\hat p_k)
\label{eq:estwce}
\ee
for some $A'>0$.
In particular, once we have proven that $(\prod_kB_k)\r_{n+m}^0(\bze(0))$ is a well defined measurable function, 
it follows that its integral in any of the variables $\bz_n,\bt_m,\bo_m,\hat\bp_m$ is absolutely convergent. We will
use this fact repeatedly during the proof of our results.



\subsection{A graphical expression of \eqref{eq:introresult}}
\label{subsec:family}

Due to the structure of the collision histories, as described in the previous section, it is quite natural 
to graphically represent each term in the expansion \eqref{eq:seriesrev} as a binary tree.
Let us introduce at a formal level the useful family of graphs. 

For fixed $n$, we define the {\em $m-$node, $n-$particle tree graph}, denoted $\TT_{n,m}$, as the collection of integers
$j_1,\cdots,j_m$ appearing in the right hand side of Eq. \eqref{eq:seriesrev}, i.e.
\be
j_1\in I_n, j_2 \in I_{n+1}, \cdots, j_m\in I_{n+m-1}\;,\ \ \ \ \ \ \mbox{with\ \ \ \ \ \ $I_k=\{1,2,\cdots,k\},$}
\ee
so that we shall write
\be
\sum_{\substack{j_1,\cdots,j_m\\ j_k\in I_{n+k-1}}} = \sum_{\TT_{n,m}}\;. \label{eq:defsumtrees}
\ee

\begin{figure}[htbp] 
   \centering
   \includegraphics[width=5 in]{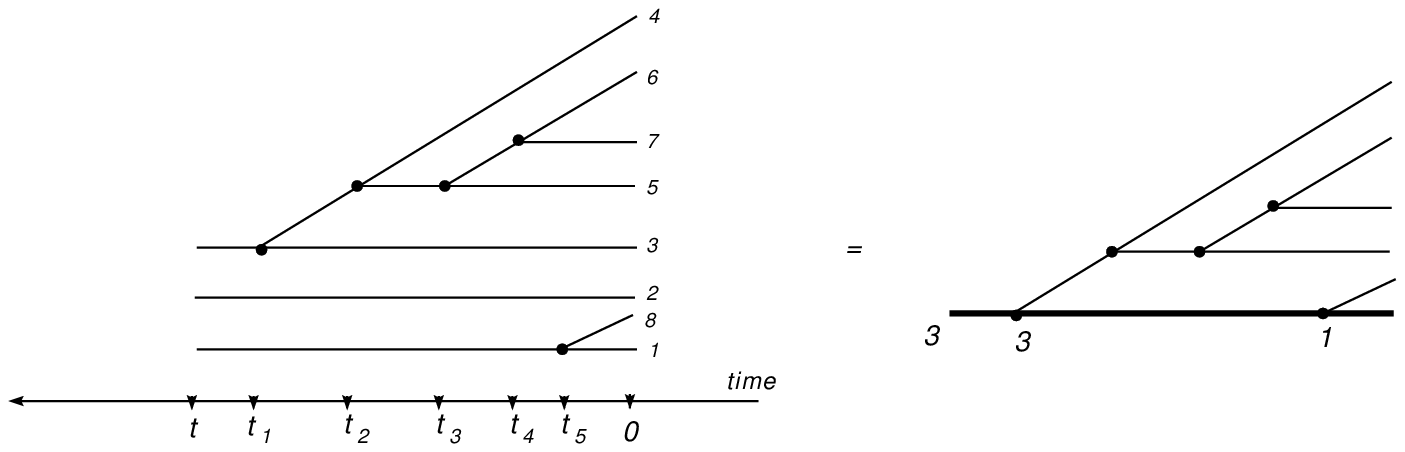} 
   \caption{Tree graph $\TT_{3,5} = 3,4,5,6,1$. On the right, a second equivalent representation.
   }
   \label{fig:example}
\end{figure}

This has an equivalent graphical representation, given by the following simple procedure. 
First, draw $n$ horizontal lines, all of them with the same length, stacked one above the other. Assign them the numbers 
$1,2,3,\cdots,$ from the bottom upwards. We will refer to such lines as the ``root lines'' of the tree graph. Time will be thought 
as flowing from right to left along a horizontal axis, in such a way that the left extremum of the segments corresponds to time 
$t$ while the right corresponds to time zero.
Now, if $m\geq 1$, draw a heavy dot over the line $j_1$ (so that the line ``crosses'' the dot) and a new straight line with a certain 
slope (say, between $0$ and $\p/2$), having left extremum in the dot and right extremum at time zero. We shall call the dot
``node $1$'' and the new added segment ``line $n+1$''. Node $1$ will correspond to a time $t_1 \in (0,t)$. If $m\geq 2$,
draw a heavy dot (``node $2$'') over the line $j_2$, corresponding to a time $t_2\in(0,t_1)$ (hence on the right with respect to 
node $1$), and a new straight line (``line $2$'') having left extremum in the dot and right extremum at time zero. This new line 
shall be horizontal if attached (through the node) to a sloped line, and sloped if attached to a horizontal line. Finally, iterate 
these operations until the last node (``$m$") and the last line (``$n+m$'') are added. We may agree to avoid intersections 
between lines. Right extrema of the lines of a tree will be called ``endpoints'', while left extrema of the root lines will be called 
``roots''. An example of tree graph is given in Figure \ref{fig:example}.

\noindent As shown in the figure, an alternative graphical representation can be given by superposing the $n$ root lines.
In this case the (only) root line of the graph is special: it will be drawn as a bold line and it will be decorated as follows: 
(i) a label $n$ is attached to the root; (ii) if node $k$ lies on the root line, a label $j_k$ is attached to it.
Though the graph in the left hand side of Figure \ref{fig:example} is perhaps more standard, we will sometimes use decorated 
trees as in the right, in order to avoid large diagrams. Furthermore, to simplify the notation, we will not add to the drawing the axis 
of time, and we will not indicate explicitly the names of the lines (see the right hand side of Figure \ref{fig:example}).

Note that two tree graphs are ``equivalent'' if they can be superposed, together with their labels and without altering their 
topological structure neither the ordering of its nodes. In other words, the nodes of a tree are ordered along the time axis,
so that the total number of different graphs $\TT_{n,m}$ is $n(n+1)\cdots(n+m-1)$.

Since a collision history is identified by the collection of variables listed on page \pageref{list:ch}, we see that $\TT_{n,m}$ 
can be associated to a {\em class} of collision histories. Namely, $\TT_{n,m}$ represents all the collision histories with $n$ 
particles at time $t$ and $m$ particles added during the backwards evolution, with progenitors specified by $j_1,\cdots,j_m.$
In this sense, we have the dynamical interpretation of graphs:

\noindent -- root lines: particles $1,\cdots,n$ of the history, leaving from time $0$ to time $t;$

\noindent -- line $n+k:$ particle $n+k$ of the history, leaving from time $0$ to time $t_k$;

\noindent -- node $k:$ binary collision in which particle $n+k$ is created from the progenitor $j_k.$

\noindent To have a precise correspondence between single collision histories and graphs it would be sufficient to add
to the picture the following decorations: (i) labels $z_1,\cdots,z_n,t$ attached to the roots of the graph indicating the
starting configuration and the time span; (ii) triples $(t_1,\o_1,\hp_1),\cdots, (t_m,\o_m,\hp_m)$
attached to the nodes $1,\cdots,m,$ specifying the times of creation of the added particles and their position
and momenta. Of course all these decorations should satisfy the constraints listed on page \pageref{list:ch}.
%
%
%

It is important to keep in mind that two particles of the collision history can interact many times during their common lifetime.
In general, any couple of particles appearing in the graph at a given time can be in a collision configuration.  The interactions 
which are {\em not} creations (and occur usually in the open time intervals $(t_{k+1},t_k)$) will be called {\em recollisions}.
In fact, they may generally involve particles that have already interacted at some creation time (in the future) with another particle 
of the history. 

We conclude this section by rewriting Eq. \eqref{eq:seriesrev} as
\be
\r_n(\bz_n,t) = \sum_{m=0}^{\infty} \sum_{\TT_{n,m}} V(\TT_{n,m})(\bz_n,t)\;,
\label{eq:expift}
\ee
where the {\em value of the tree} $V(\TT_{n,m})$ is
\be
V(\TT_{n,m})(\bz_n,t) =\int_{\CC_{\TT_{n,m}}(\bz_n,t)} d\m(\bt_m,\bo_m,\hat\bp_m) \left(\prod_{k=1}^m 
B(\o_k; \hp_k-\p_{j_k}(t_k))\right)\r_{n+m}^0(\bze(0))\;, \label{eq:valtr}
\ee
i.e. the integral of the initial datum $\r_{n+m}^0$, with a suitable weight, over all the possible time--zero states 
of the collision histories associated to $\TT_{n,m}$. In the next section, to represent graphically $V(\TT_{n,m})(\bz_n,t)$, 
we will just draw the graph $\TT_{n,m}$ as in Figure \ref{fig:example} and attach to the root a label $\bz_n$.

\section{The evolution of correlation functions} \label{sec:res}

In what follows we present our main theorem (Theorem \ref{thm:result}). Then, we
derive the usual BBGKY hierarchy of equations (Corollary \ref{cor:resultB}). Finally, we present an extension of the result to 
measures of grand canonical type (Corollary \ref{cor:result}).

\begin{thm} \label{thm:result}
Given an initial measure on $\G_N$ with density $f_N^0 \in\mathcal L_N$, let $f_N(t)$ be the time--evolved density and 
$\r_n(t)$ the associated correlation functions, as defined respectively in \eqref{eq:liouville} and \eqref{eq:defcf}.
Then, the expansion \eqref{eq:expift}--\eqref{eq:valtr} holds for any $t>0$, almost everywhere in $\G_n$. If
\eqref{eq:liouville} and \eqref{eq:defcf} are satisfied over the whole sets $\G_n^{\dagger},$ then the expansion is valid 
for all $(\bz_n,t) \in \G_n^{\dagger}\times\RRR^+$.
\end{thm}



As mentioned in the introduction, unlike in \cite{Spohn} and in \cite{IP}, we do not need $f_N^0$ and $\r_n^0$ to be 
``continuous along trajectories'', that is we do not need
\be
\lim_{s\rightarrow 0} f_N^0(T^{(N)}_s(z_1,\cdots,z_N)) = f_N^0(z_1,\cdots,z_N) \label{eq:conttraj}
\ee
for a.a. $\bz_N\in\G_N,$ where both the limits from the future and the past are understood.
If the continuity along trajectories is assumed to be valid for $f_N^0$, then the Liouville Equation \eqref{eq:liouville} together 
with some integrability bound on $f_N^0$ imply that the same continuity property holds for $f_N(t)$ and for $\r_n(t)$ at any 
time $t \geq 0$, and that the map $t\rightarrow\r_n(\bz_n,t)$ is also continuous for almost all $\bz_n$ (see \cite{Spohn},
where this is proved and used). All these properties, even if assumed, would be not helpful in the proof of Section \ref{sec:proof}.

As for the control on large momenta, assumption \eqref{eq:assbound} could be substituted with a weaker one, 
since it will be actually needed just to ensure the absolute convergence of the integrals in the expansion: see 
\eqref{eq:estwce} and the comment therein.
Our choice of the decay behaviour for high momenta is the same used by Lanford in the
careful estimates of \cite{Lanford} (see the details in \cite{King}), necessary to perform the Boltzmann--Grad limit.

Finally, observe that our result is actually stronger than the one obtained in the previous literature \cite{Spohn}, \cite{IP},
\cite{Uchiyama}. We know by \cite{Alexander} that there exists a full measure subset of the phase space where the dynamics 
of the hard sphere system exists for all times (Proposition \ref{prop:dyn}). (The last statement of) Theorem \ref{thm:result} 
recovers this property for the evolution of correlation functions. Unfortunately the subset $\G_n^{\dagger}$, in which the expansion 
for the correlations is valid for all times, has not been characterized in a constructive manner (see Proposition~\ref{prop:dynbis}): 
this would depend on details of the dynamics that have not been investigated. However, it will be clear from the proof, which method
{\em is} instead constructive, that Eq. \eqref{eq:Gndagmax} defines the maximal subset of the phase space where the result can 
be derived for all times, as soon as $\G_n^*$ is given as the maximal subset on which the hard sphere dynamics is well defined.
In particular, the last statement of our theorem will be still true if we replace $\G_n^{\dagger}$ with any full 
measure invariant subset of it, say $\HH_n,$ satisfying the following ``chain property'': if $\bz_n\in\HH_n,$ then 
$(\bz_n,\by_k)\in\HH_{n+k}$ for almost all $\by_k \in \G_k(\bz_n).$

\vspace{2mm}
Let us turn now to the usual BBGKY hierarchy of integro--differential equations. The hierarchy can be recovered, though in 
a mild sense, from the expansion \eqref{eq:expift}. 

The {\em collision operator} $Q$ acting on the time--evolved correlation function (abusing the notation used in the introduction)
is defined by
\be
\left(Q \r_{n+1}\right) (\bz_n, t) = a^2 \sum_{j=1}^n \int_{\RRR^3 \times \O_{j}(\bz_n)}d\hat p\ d\o\ \o\cdot \left(\hat p -p_j\right)
\r_{n+1}\left(\bz_n, q_j+ a\o,\hat p, t\right)\;. \label{eq:Qdef}
\ee
%
This definition does not depend (almost surely) on values assumed by the initial measure on a set of measure zero.
Suppose indeed that $f_N(0),\tilde f_N(0) \in \LL_N$, with $f_N(0) = \tilde f_N(0)$ a.s. in $\G_N$. Of course the 
Liouville equation implies that this 
remains true for any positive time. But, by the property in Remark 2 of page \pageref{rem:dyncs}, the same is true also for
almost all $(\bz_N,t)\in\partial\G_N\times\RRR$. This implies $\r_n(t)=\tilde\r_n(t)$ for a.a. $(\bz_n,t)\in\partial\G_n\times\RRR$,
so that $Q\r_{n+1}=Q\tilde\r_{n+1}$ a.s. in $\G_n\times\RRR$. 

It holds
\begin{cor} \label{cor:resultB}
Given an initial measure on $\G_N$ with density $f_N^0 \in\mathcal L_N$, let $f_N(t), \r_n(t)$ satisfy \eqref{eq:liouville},
\eqref{eq:defcf} over $\G_n^{\dagger}$. Then the function $t \longrightarrow \left(Q\r_{n+1}\right) (T^{(n)}_{t}\bz_n,t)$ 
is $dt-$measurable and $t \longrightarrow \r_n(T^{(n)}_{t}\bz_n,t)$ is absolutely continuous, for all $\bz_n\in\G_n^{\dagger}$.
The correlation functions satisfy
\be
\frac{d}{dt}\r_n(T^{(n)}_{t}\bz_n,t) = \left(Q\r_{n+1}\right) (T^{(n)}_t\bz_n,t)
\label{eq:BBGKY}
\ee
for all $\bz_n\in\G_n^{\dagger}$ and almost all $t > 0$.
\end{cor}

\noindent The result, which strengthens the analogous in \cite{CIP}, is obtained by resummation of the series validated 
in Theorem \ref{thm:result} (see Section \ref{sec:resultB}).

We stress that the mild continuity property stated in the corollary is a consequence of the only Liouville equation,
and it does not imply the stronger continuity--along--trajectories of the correlation functions, which is in general not 
valid unless we assume Eq. \eqref{eq:conttraj} for the initial measure.

To gain regularity in the right hand side of the hierarchy, we need further assumptions. For instance, it can be checked
that continuity in $t$ of $\left(Q\r_{n+1}\right) (T^{(n)}_t\bz_n,t)$ follows if the continuity--along--trajectories of $\r_{n+1}(t)$
holds for a.a. values of the integration variables $\hat p,\o$. This would be in turn ensured (at least for a.a. $\bz_n, t$) by 
assumption \eqref{eq:conttraj}, or also by the continuity of the initial density in a full measure subset of the phase space.
We shall not pursue this further here.

\vspace{2mm}
It is worth to say that the proof of Theorem \ref{thm:result} extends easily to a more general class of measures 
with non definite (but finite) number of particles. Consider the grand canonical phase space
\be
\G = \cup_{n\geq 0}\G_n\;.
\ee
There holds $\G_n = \emptyset$ for $n$ larger then $[3|\L|/4\pi a^3]$, because of the hard core exclusion.

Call $\mathcal L$ the space of vectors of functions $\bof:\G\rightarrow\RRR, \bof = \{f_n\}_{n\geq 0}$, with $f_n\in\mathcal L_n$.
If $P$ denotes a measure on $\G$ with density $\bof^0 \in\mathcal L$ with respect to the Lebesgue measure,
then the time--evolved measure at time $t$ has a density $\bof(t)\in\mathcal L$ given by
\be
f_n(\bz_n,t) = f_n^0(T^{(n)}_{-t}(\bz_n))\;,\ \ \ \ \ \ n\geq 0 \label{eq:liouvillebis}
\ee
almost everywhere in $\G_n$.

We define the correlation function vector $\br(t) : \G\longrightarrow \RRR, 
\br = \{\r_n\}_{n \geq 0}$, by
\bea
\r_n(\bz_n,t)=\sum_{k=0}^{\infty} \frac{1}{k!} \int_{\G_{k}(\bz_n)}dz_{n+1}\cdots dz_{n+k} 
f_{n+k}(\bz_{n+k},t)\;.\label{eq:defcftris}
\eea
It is easy to check that $\br(t)\in\mathcal L$ and that, furthermore, the map defined by \eqref{eq:defcftris} has the inverse
\bea
f_n(\bz_n,t)
=\sum_{k=0}^{\infty} \frac{(-1)^k}{k!} \int_{\G_{k}(\bz_n)}dz_{n+1}\cdots dz_{n+k} 
\r_{n+k}(\bz_{n+k},t)\;.
\eea

We have the following
\begin{cor} \label{cor:result}
Given an initial measure on $\G$ with density $\bof^0 \in\mathcal L$, let $\bof(t)$ be the time--evolved density and 
$\br(t)$ the associated correlation functions, as defined respectively in \eqref{eq:liouvillebis} and \eqref{eq:defcftris}.
Then, the expansion \eqref{eq:expift}--\eqref{eq:valtr} holds for any $t>0$, almost everywhere in $\G_n$. If
\eqref{eq:liouville} and \eqref{eq:defcf} are satisfied over the whole sets $\G_n^{\dagger},$ then the expansion is valid
for all $(\bz_n,t) \in \G_n^{\dagger}\times\RRR^+$, and the results of Corollary \ref{cor:resultB} hold.

\end{cor}

\noindent Here $\G_n^{\dagger}$ is defined as in Proposition \ref{prop:dynbis}, with $k\geq 1$.
The (trivial) modifications of the proof of the main theorem leading to Corollary \ref{cor:result} will be discussed in
Section \ref{sec:resultN}.

\section{Proofs} \label{sec:proof}

To prove Theorem \ref{thm:result}, we shall proceed by induction on $n$: supposing
the claim true for the function $\r_{n+1}$, we derive the expansion for the $\r_n$
by integrating out the state of a single selected particle. The proof is organised as follows.
In Section \ref{subsec:idf} we describe the generic step of the induction. In Proposition \ref{lem:int} 
we explain what is the result when one integrates out the one--particle state in a given term (tree) of the expansion.
The proof of Proposition \ref{lem:int}, which is our main task, is discussed in Section \ref{ss:pPli}. After that,
to conclude the proof of the main theorem we have to sum the result over all possible trees, which is done in 
Section \ref{ss:sum}. Finally, in the last two sections we prove Corollaries \ref{cor:resultB}, \ref{cor:result}.

The iterative integration rule and the technical steps of Sections \ref{ss:pPli},\ref{ss:sum}, admit a quite simple graphical 
representation in terms of manipulations of tree graphs. This may help the reader to understand quickly the notations
introduced along the proof. 

The analytical operations leading to Proposition \ref{lem:int} consist in appropriate 
partitioning of the integration domain, and representation of its subsets via suitable changes of variables. Such 
parametrizations turn out to be rather simple, since they are constructed using only non--interacting one--particle trajectories.
Nevertheless, as mentioned in the introduction, this is not enough: to prove the proposition it is also essential to notice that a certain
class of collision histories gives a net null contribution to the integral, because of one by one cancellations.
This will be the content of Lemma \ref{lem:int2}. 

\subsection{Integration of a particle state} \label{subsec:idf}

For $n=N$ (and of course $n>N$) the statement of Theorem \ref{thm:result} is trivially implied by \eqref{eq:defcf} and \eqref{eq:liouville}.
Formula \eqref{eq:expift} gives
\be
\r_N(\bz_N,t) = \ \ _{\bz_N} \frac{\ \ \ \ \ \ \ \ \ \ \ \ \ \ \ }{\ \ \ \ \ \ \ \ \ \ \ \ \ \ \ } = \r_N^0(T^{(N)}_{-t}\bz_N)\;.
\ee

We proceed by induction on $n$. From \eqref{eq:defcf} it follows
\be
\r_n(\bz_n,t) = \frac{1}{N-n}\int_{\G_1(\bz_n)}dz_{n+1} \r_{n+1}(\bz_n,z_{n+1},t)\;,\ \ \ \ \ \ \ \ 1\leq n<N\;.
\label{eq:defcralt}
\ee
Let us assume that, for any $t>0$, $V(\TT_{n+1,m})$ is a Borel function over $\G_{n+1}$ with absolute value bounded by 
$A'\prod_{j=1}^{n+1} h_{\b'}(p_j)$, for some $A',\b'>0$, and that Eq. \eqref{eq:expift} is valid for $\r_{n+1}$.
Then we can write
\bea
&& \r_n(\bz_n,t) =  \frac{1}{N-n}\sum_{m=0}^{\infty} \sum_{\TT_{n+1,m}}\II(\TT_{n+1,m})(\bz_n,t)\;,\label{eq:sIIT}\\
&& \II(\TT_{n+1,m})(\bz_n,t)=\int_{\G_1(\bz_n)}dz_{n+1}V(\TT_{n+1,m})(\bz_{n+1},t)\;, \label{eq:IIT}
\label{eq:rnesIT}
\eea
a.e. in $\G_n$. 

The two last equations hold exactly in $\G_n^{\dagger}$ if \eqref{eq:liouville} and \eqref{eq:defcf} are satisfied over the 
corresponding spaces. Unless where explicitly stated, we may assume that this is true from now on: $f_N,\r_n$ satisfy 
\eqref{eq:liouville}, \eqref{eq:defcf} over the whole sets $\G_n^{\dagger}$, and we fix $(\bz_n,t)\in\G_n^{\dagger}\times\RRR^+$.
If this is not the case, it will be clear that each step of the proof that follows is still valid in some full measure, possibly 
$t-$dependent, subset of $\G_n$.

In the rest of this section and in the next one, we will focus on the computation of \eqref{eq:rnesIT}.

\subsubsection*{Integration of a particle state in a single tree}

Let us explain what is the result when we integrate a particle state in a given tree.
The computation of \eqref{eq:IIT} will be the main part of the proof, and the content of Section \ref{ss:pPli}.

The bulk of Theorem \ref{thm:result} is contained in the following assertion.
\begin{prop} \label{lem:int}
Fix $n,m,\TT=\TT_{n+1,m} = j_1,\cdots, j_m$. Let $\ell=m+1$ if $\{k\ |\ j_k=n+1\}=\emptyset$, and $\ell = \min\{k\ |\ j_k=n+1\}$ otherwise.
There holds
\be
\II(\TT_{n+1,m})= \d_{\ell,m+1}\left( N-n -m \right) V(\TT''_{n,m}) +
\sum_{k = 1}^{\ell}\sum_{i = 1}^{n+k-1}V(\TT'_{n,m+1})\;, \label{eq:ITrule}
\ee
where $\TT''=\TT''_{n,m}=\bj''_m$ and $\TT'=\TT'_{n,m+1}=\bj'_{m+1}$ are the $n-$particle trees given by the rules
\bea
&& \bj''_m = f''(j_1),\cdots,f''(j_m) \nn\\
&& f''(j) = 
\left\{
\begin{array}{cc}
j  &  \mbox{ \ \ \ if \ \ } j\leq n   \\
j-1 & \mbox{ \ \ \ \ \ \ \ \ if \ \ }  j \geq n+2    \\
\end{array}
\right. \label{eq:TTs}
\eea
and
\bea
&& \bj'_{m+1} = f'(j_1),\cdots,f'(j_{k-1}),i,f'(j_k),\cdots,f'(j_m)\nn\\
&& f'(j) = 
\left\{
\begin{array}{cc}
j  &  \mbox{ \ \ \ \ \ \ \ \ \ \ \ \ if \ \ } j\leq n, j\geq n+k+1   \\
j-1 & \mbox{ \ \ \ \ \ \ \ \ \ \ if \ \ } n+2 \leq j \leq n+k    \\
n+k & \mbox{ if \ \ } j=n+1
\end{array}
\right.\;. \label{eq:TTp}
\eea
All the terms on the r.h.s. of \eqref{eq:ITrule} are Borel functions over $\G_n$ with absolute value bounded by 
$A'\prod_{j=1}^n h_{\b'}(p_j)$, for some $A',\b'>0$.

\end{prop}
\noindent Here $\d$ indicates the Kronecker delta. Notice that we drop the dependence on $k,i$ of the trees $\TT'$.

\begin{figure}[htbp] 
   \centering
   \includegraphics[width=5in]{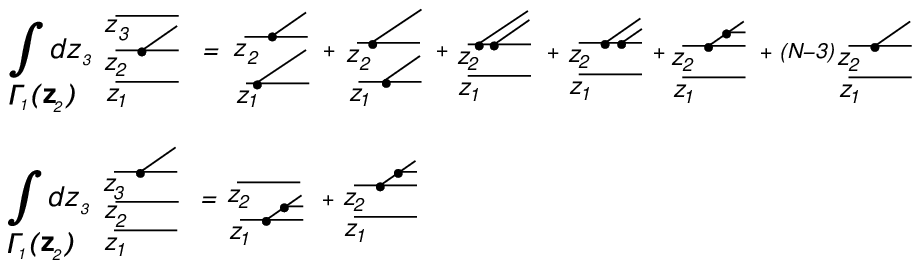} 
   \caption{Computation of the integral $\II$ for two tree graphs $\TT_{n+1,m}$ with $n=2, m=1$. In the first line, a case with $\ell=2$.
   In the second line, a case with $\ell=1$. Notice that in the first line, on the right hand side, the third and the fourth graphs are equivalent,
   while the last graph is produced by operation 3 of the list above.}
   \label{fig:proofnd}
\end{figure}

Representing the tree graph $\II(\TT_{n+1,m})$ as made of $n+1$ distinct trees, as in the left hand side of Figure~\ref{fig:example}, 
we can give the following picture of Proposition \ref{lem:int}. To compute $\II(\TT_{n+1,m})$:
\begin{enumerate} \label{list:rules}
\item Consider the $(n+1)-$th tree graph in $\TT_{n+1,m}$, i.e. the tree having line $n+1$ as root (note that $\ell$ is defined as the 
name of the first node of this tree, if any, going from left to right). {\em Attach} its root to the line $i$ of 
$\TT_{n+1,m}$, between node $k-1$ and node $k$, taking care to {\em preserve} the reciprocal ordering of the nodes of 
$\TT_{n+1,m}$. The (only possible) resulting tree, $\TT'_{n,m+1}$,
will have the old $m$ nodes of $\TT_{n+1,m}$, plus one {\em new} node coming from
this last operation. Compute now the value of the resulting tree.
\item Sum the result of the previous point over all possible choices of $k$ and $i$.
\item If the $(n+1)-$th tree graph in $\TT_{n+1,m}$ is trivial (i.e. it has no nodes), add to the result of point 2 the value of the 
$n-$particle tree obtained by {\em discarding} the trivial line, i.e. $\TT''_{n,m}$, multiplied by a factor $(N-n-m)$.
\end{enumerate}
See Figure \ref{fig:proofnd} for an example. 
\begin{figure}[htbp] 
   \centering
   \includegraphics[width=5in]{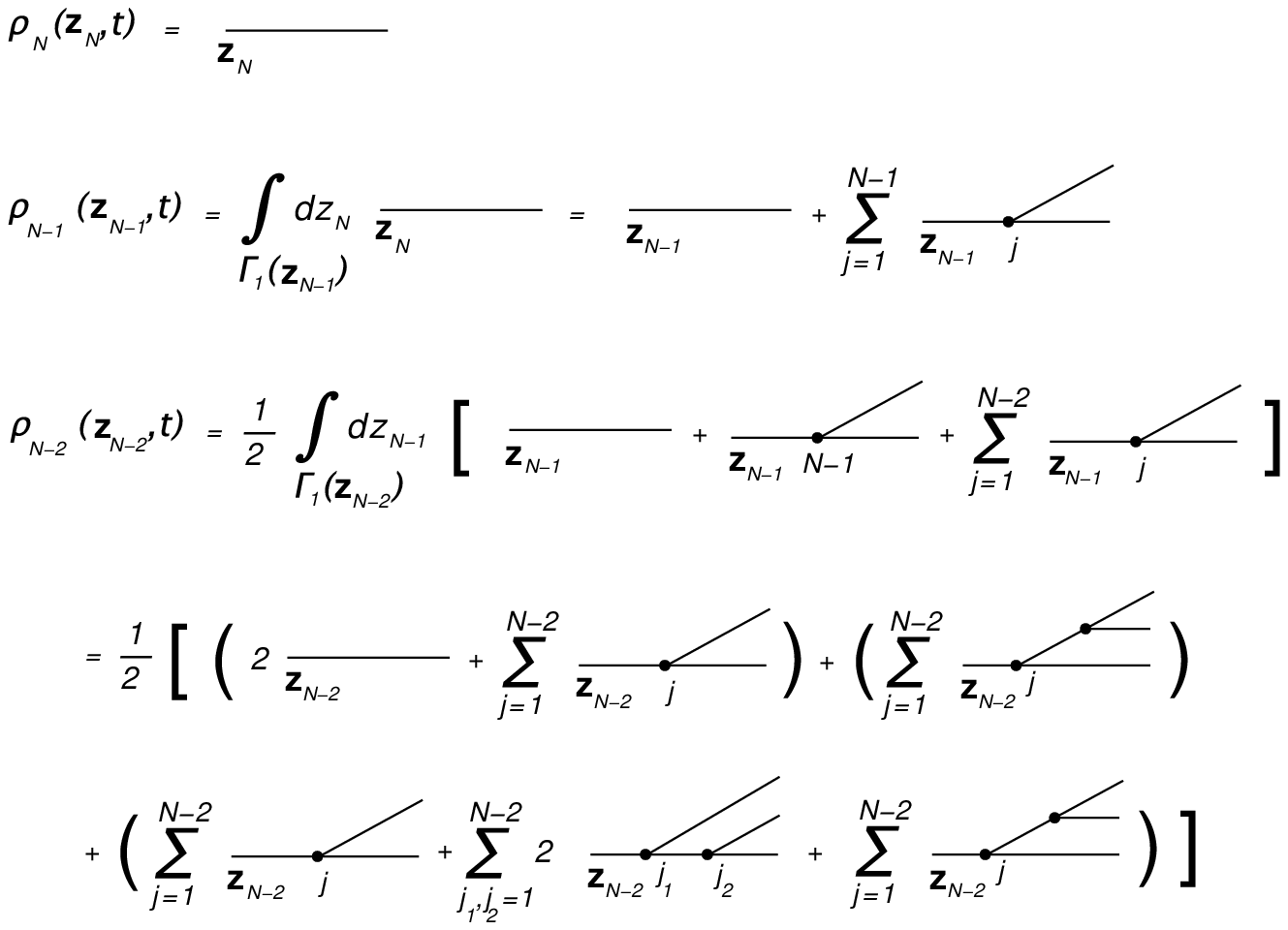} 
   \end{figure}
\begin{figure}[htbp] 
   \centering
   \includegraphics[width=6in]{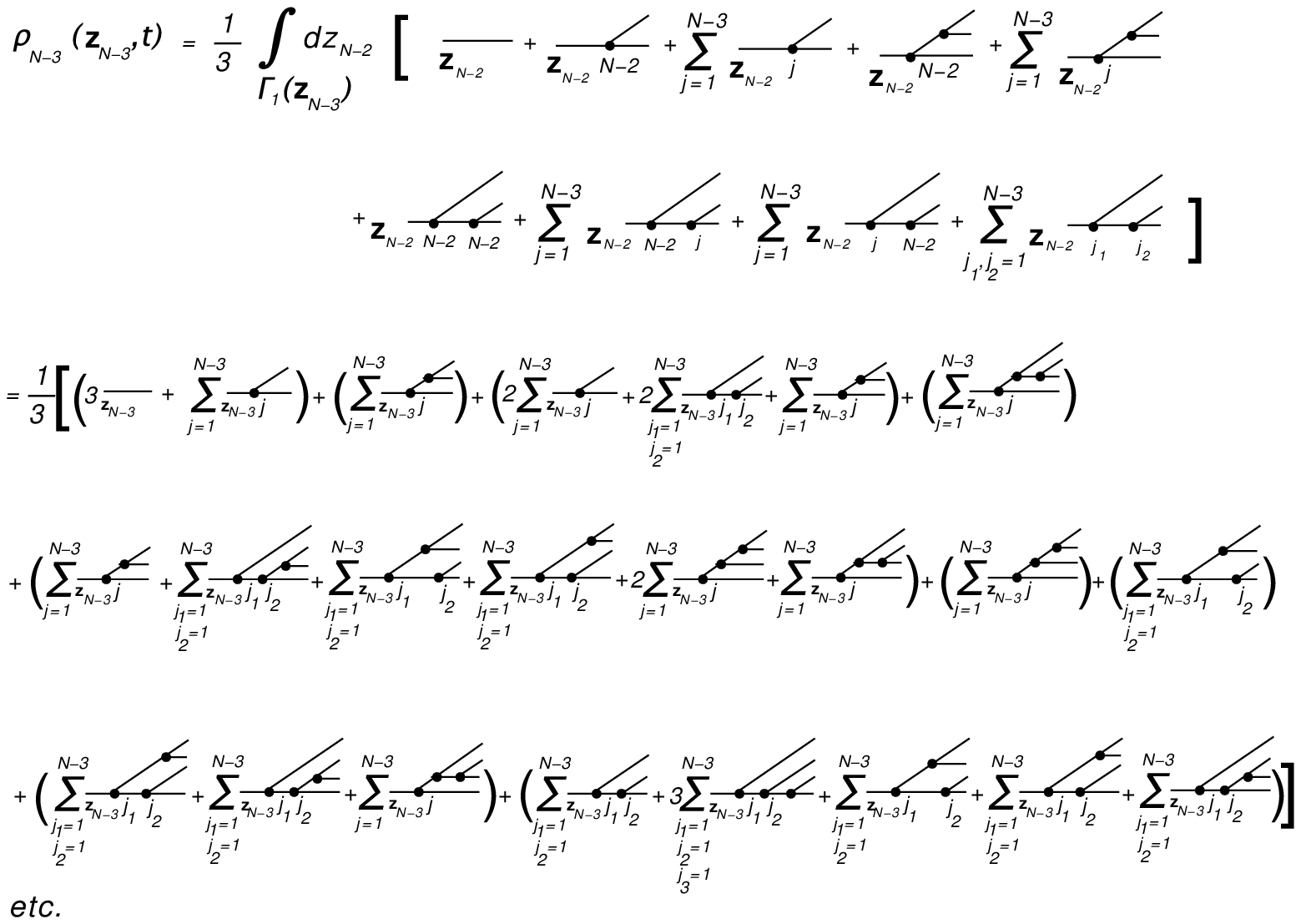} 
   \caption{Integration of degrees of freedom: from Liouville equation to BBGKY hierarchy.}
   \label{fig:proof}
\end{figure}
Several other examples are provided by Figure \ref{fig:proof}, in which the 
alternative graphical representation introduced on the right hand side of Figure \ref{fig:example} is used in order to avoid 
too large diagrams.

Observe that the new node described in point $1$ of the list has number $k \leq \ell$ in the resulting tree $\TT'_{n,m+1}$, while the other 
nodes have to be consequently renamed: those on its left conserve their name, while those on its right increase of a unit. 
In particular, if $\ell < m+1$, in the resulting tree the node $\ell + 1$ is the first one ``crossed'' by line $n+k$.

\vspace{2mm}
We close this section by giving an idea of how formula \eqref{eq:ITrule} emerges. To do so, let us discuss (briefly and somewhat loosely)
the first nontrivial step, namely $n = N-1$, or second line in Figure \ref{fig:proof}. In this case, $m$ has to be $0$ (and $\ell =1$), and we only 
need to compute $\int_{\G_1(\bz_{N-1})}dz \ \r_N^0(T^{(N)}_{-t}(\bz_{N-1},z))$.

Consider the backwards trajectory leading from $(\bz_{N-1},z)$ at time $t$,
to $T^{(N)}_{-t}(\bz_{N-1},z)$ at time $0$. Either the last particle (``particle $N$") goes freely, or interacts with one of the other $N-1$ 
particles. We make accordingly the partition $\G_1(\bz_{N-1}) = \G'_0 \cup {\G'_0}^c$, ${\G'_0}^c = \cup_{j_1=1}^{N-1} \G'_{j_1}$, with 
$j_1 = $ index of the first particle encountered by particle $N$ in its backwards motion. First, we reexpress $\int_{\G'_{j_1}}dz$ through
the change of variables $z \equiv (q,p) \to (t_1,\o_1,\hat p_1)$, where $t_1$ is the time of the first (backwards) interaction between the particles
$N$ and $j_1$, $q - p(t-t_1) = q_{j_1}(t_1) + a \o_1$, and $\hat p_1 = p$. Here $q_{j_1}(t_1), p_{j_1}(t_1)$ are position and momentum of
particle $j_1$ at time $t_1$, evolved with the $(N-1)-$particle dynamics. The volume element transforms as 
$dz = a^2\o_1 \cdot (\hp_1- p_{j_1}(t_1)) dt_1 d\o_1 d\hp_1$. That is, using the notations of \eqref{eq:seriesrev}--\eqref{eq:notterm}, 
\be
\int_{\G_1(\bz_{N-1})}dz \ \r_N^0(T^{(N)}_{-t}(\bz_{N-1},z)) = \int_{\G'_0}dz \ \r_N^0(T^{(N)}_{-t}(\bz_{N-1},z))
+ \sum_{j_1=1}^{N-1} \int_+ d\m_1\ \mathbbm{1}_{B_1>0}\ B_1\ \r_N^0(\bze(0))\;,
\ee
where $\int_+$ is restricted to trajectories such that particle $N$ moves freely in the time interval $(t_1,t)$.

Next, we observe that, if $\ol z = (q-pt,p)$, then $\int_{\G'_0}dz \ \r_N^0(T^{(N)}_{-t}(\bz_{N-1},z)) = \int_{\G''_0}d\ol z \
\r_N^0(T^{(N-1)}_{-t}\bz_{N-1},\ol z)$, being $\G''_0$ the subset of $\G_1(T^{(N-1)}_{-t}\bz_{N-1})$ such that particle $N$ moves freely in 
the time interval $(0,t)$. To this result
we add and subtract the integral over ${\G''_0}^c = \G_1(T^{(N-1)}_{-t}\bz_{N-1}) \setminus \G_0'' = \cup_{j_1=1}^{N-1} \G''_{j_1}$,
with $j_1 = $ index of the first particle encountered by particle $N$ in its forward motion.
We find $\r_{N-1}^0(T^{(N-1)}_{-t}\bz_{N-1}) - \sum_{j_1=1}^{N-1} \int_{\G''_{j_1}}d\ol z \ \r_N^0(T^{(N-1)}_{-t}\bz_{N-1},\ol z)$. Now we proceed as before,
i.e. we change variables according to $\ol z \equiv (\ol q,\ol p) \to (t_1,\o_1,\hat p_1)$, where $t_1$ is the time of the first (forward) interaction between 
particles $N$ and $j_1$, $\ol q + \ol p t_1 = q_{j_1}(t_1) + a \o_1$, and $\hat p_1 = \ol p$. The final result is
\bea
&& \int_{\G_1(\bz_{N-1})}dz \ \r_N^0(T^{(N)}_{-t}(\bz_{N-1},z)) \\
&& = \r_{N-1}^0(T^{(N-1)}_{-t}\bz_{N-1}) + \sum_{j_1=1}^{N-1} \int_+ d\m_1\ \mathbbm{1}_{B_1>0}\ B_1\ \r_N^0(\bze(0))
- \sum_{j_1=1}^{N-1} \int_- d\m_1\ \mathbbm{1}_{B_1<0}\ |B_1|\ \r_N^0(\bze(0))\;. \nn
\eea

\begin{figure}[htbp] 
   \centering
   \includegraphics[width=6.5in]{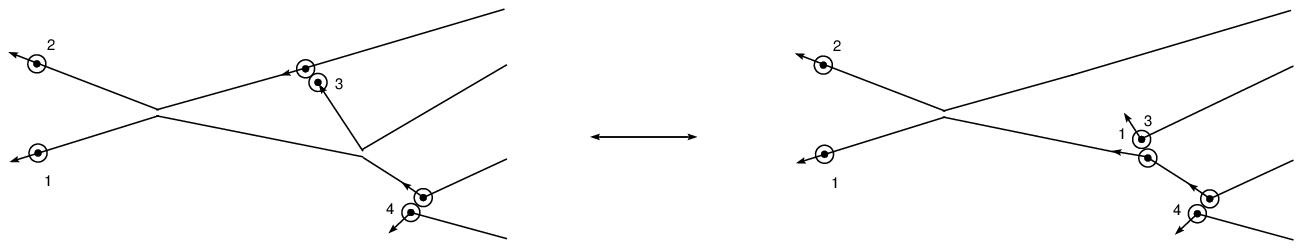} 
   \caption{Cancellations between collision histories. A case with $k=1, i=2, k^*=1, i^*=1$.}
   \label{fig:csJSP}
\end{figure}

To reconstruct the left hand side of \eqref{eq:ITrule}, we need finally to get rid of the restrictions $+ / -$ under the signs of integral.
Call respectively $+^c / -^c$ the complements of these restrictions. These are values of $(t_1,\o_1,\hat p_1)$ such that particle $N$
undergoes a collision in the forward / backwards evolution starting from $(\xi_{j_1}(t_1) + a \o_1,\hp_1)$. A one--to--one mapping
is naturally established between $+^c$ and $-^c$, by looking at the first forward / backwards collision starting from $(\xi_{j_1}(t_1) + a \o_1,\hp_1)$
(see e.g. Figure \ref{fig:csJSP} on page \pageref{fig:csJSP}). For instance, we may rewrite the $\sum_{j_1=1}^{N-1}\int_{-^c}d\m_1$ by applying 
the transformation $(j_1, t_1,\o_1,\hat p_1) \to (j'_1, t'_1,\o'_1,\hat p'_1)$, where $t'_1$ is the time of the first (backwards) interaction of particle 
$N$ in $(0,t_1)$, $j'_1$ is the index of the particle involved in such collision, $\o'_1$ is the unit vector indicating the relative position of $N$ with 
respect to $j'_1$ at time $t'_1$, and $\hat p_1 = \hat p'_1$. Since the volume element transforms as $- a^2 \o_1 \cdot (\hat p_1-p_{j_1}(t_1)) dt_1 d\o_1
d\hat p_1 = a^2 \o'_1 \cdot (\hat p'_1-p_{j'_1}(t'_1)) dt'_1 d\o'_1 d\hat p'_1$, we obtain that 
$\sum_{j_1=1}^{N-1} \int_{-^c} d\m_1\mathbbm{1}_{B_1>0}B_1\r_N^0(\bze(0)) =
\sum_{j_1=1}^{N-1} \int_{+^c} d\m_1\mathbbm{1}_{B_1<0}|B_1|\r_N^0(\bze(0))$. This concludes the proof of second line in Figure \ref{fig:proof}.

\subsection{Proof of Proposition \ref{lem:int}} \label{ss:pPli}

Before starting the proof, we need some additional notation. First of all, in this section we shall drop the lower indices
in the names of the trees, unless where stated, and use the symbols $\TT,\TT',\TT''$ introduced by Proposition \ref{lem:int}. 
To avoid confusion, we will mark with a symbol $'$ (or~$''$) the variables of the collision histories associated to $\TT'$ 
(or $\TT''$) of Proposition~\ref{lem:int}, and without that symbol those associated to the tree $\TT$. 
More precisely, if the variables
\be
\bz_{n+1}, t, j_1,\cdots,j_m,t_1,\cdots,t_m,\hat p_1,\cdots,\hat p_m,\o_1,\cdots,\o_m
\ee
describe the collision histories $\bze$ associated to $\TT$, then
\be
\bz_n, t, j'_1,\cdots,j'_{m+1},t'_1,\cdots,t'_{m+1},\hat p'_1,\cdots,\hat p'_{m+1},\o'_1,\cdots,\o'_{m+1} \label{eq:chprime}
\ee
describe the collision histories $\bze'$ associated to $\TT'$ (where the $\bj'_{m+1}$ are given by \eqref{eq:TTp}).
A similar notation will be used for $\TT''$.
We recall also the notations $t_0=t=t'_0=t''_0$, $t_{m+1}=0=t'_{m+2}=t''_{m+1}$, that will be used in the sequel.

For generic $\bze = \z_1,\z_2,\cdots$, with $\z_i=(\xi_i,\p_i)$ and $z = (q,p)$, we put 
\be
\dist(\bze(s),z) = \min_i|\xi_i(s)-q|\;,
\ee
i.e. the minimum distance, in position space, of a particle in $z$ from the cluster of particles of the collision history at time $s$.
Similarly, we put
\be
\dist_k(\bze(s)) = \min_{i\neq k}|\xi_i(s)-\xi_k(s)|\;,
\ee
that is the minimum distance of particle $k$ of the history from the other particles of the same history at time~$s$.

When we need to specify positions and momenta of a generic configuration $z_1,\cdots,z_n$, with $z_i=(q_i,p_i)$, 
evolved at time $s$ with the $n-$particle dynamics, we shall use the notation
\be
(q^{(n)}_j(s),p^{(n)}_j(s))\;,\ \ \ \ \ j=1,\cdots,n\;.
\ee

Let us introduce (for the moment formally) some special subsets of the integration domains in computing the value 
of $\TT'$. Call
\bea
&& \FF^+_{k,i} = \Big\{ (\bt'_{m+1},\bo'_{m+1},\hat\bp'_{m+1}) \mbox{\ \ s.t.\ \ } \o'_k\cdot(\hat p'_k-\p'_{j'_k}(t'_k))>0 \mbox{ \ and\  }\nn\\
&& \dist\left(\bze'(s),T^{(1)}_{-t'_k+s}\left(\xi'_{j'_k}(t'_k)+a\o'_k,\hp'_{k}\right)\right) > a \mbox{ for all }s\in(t'_k,t) \Big\}\;,\nn\\
&& \FF^-_{k,i} = \Big\{ (\bt'_{m+1},\bo'_{m+1},\hat\bp'_{m+1}) \mbox{\ \ s.t.\ \ } \o'_k\cdot(\hat p'_k-\p'_{j'_k}(t'_k))<0 \mbox{ \ and\  }\nn\\
&& \dist_{n+k}\left(\bze'(s)\right) > a \mbox{ for all }s\in(t'_{\ell+1},t'_k) \Big\}\;.
\eea
In other words, $\FF^+_{k,i}$ selects those collision histories associated to $\TT'$ which satisfy the special property explained 
as follows. Consider particle $n+k$ of the collision history $\bze'$, i.e. the particle created in the ``new'' node of $\TT'$
Assume that this particle is created in an outgoing collision 
configuration. Its state at the moment of creation is $(\xi'_{j'_k}(t'_k)+a\o'_k,\hp'_{k})$. Then, if we evolve forward in time such a
state up to time $t$, we do not see any interaction of the particle with any of the hard spheres appearing in the evolution $\bze'$. 
Similarly in $\FF^-_{k,i}$, if we evolve particle $n+k$ (created in an incoming collision configuration) backwards in time up to the 
time in which it creates another particle of the history (if any; or up to zero otherwise), then we do not see any interaction of it with the 
hard spheres appearing in the evolution $\bze'$.

Abbreviating here $z^{(1)}_{n+k}(s)=(q^{(1)}_{n+k}(s),p^{(1)}_{n+k}(s)) = T^{(1)}_{-t'_{k}+s}(\xi'_{j'_{k}}(t'_{k})+a\o'_{k},\hp'_{k})$,
we shall complement the above definitions with
\bea
&& \RR^+_{k,i} = \Big\{ (\bt'_{m+1},\bo'_{m+1},\hat\bp'_{m+1}) \mbox{\ \ s.t.\ \ } \o'_k\cdot(\hat p'_k-\p'_{j'_k}(t'_k))>0 \mbox{ \ and\  
$\exists\ i_+$ and $s_+\in(t'_{k},t),\ \ $}\nn\\
&& \dist\left(\bze'(s),z^{(1)}_{n+k}(s)\right) > a \ \ \forall s\in(t'_{k},s_+),\ \ \Big|q^{(1)}_{n+k}(s_+)-\xi'_{i_+}(s_+)\Big| = a\;,\nn\\
&& \left(q^{(1)}_{n+k}(s_+)-\xi'_{i_+}(s_+)\right)\cdot\left(p^{(1)}_{n+k}(s_+)-\p'_{i_+}(s_+)\right)<0\Big\}\;, \nn\\
&& \RR^-_{k,i} = \Big\{ (\bt'_{m+1},\bo'_{m+1},\hat\bp'_{m+1}) \mbox{\ \ s.t.\ \ } \o'_k\cdot(\hat p'_k-\p'_{j'_k}(t'_k))<0  
\mbox{ \ and\ $\exists\ i_-$ and $s_-\in(t'_{\ell+1},t'_k),\ \ $}\nn\\
&& \dist_{n+k}\left(\bze'(s)\right) > a \ \ \forall s\in(s_-,t'_k),\ \ \Big|\xi'_{n+k}(s_-)-\xi'_{i_-}(s_-)\Big| = a\;,\nn\\
&& \left(\xi'_{n+k}(s_-)-\xi'_{i_-}(s_-)\right)\cdot\left(\p'_{n+k}(s_-)-\p'_i(i_-)\right)>0\Big\}\;. \label{eq:defR+R-}
\eea

Finally, we denote the restriction of the integral defining $V(\TT')$ to any subset $A$ of the integration region as
\bea
&& V |_A(\TT')=\int_{\CC_{\TT'}(\bz_n,t)} d\m(\bt'_{m+1},\bo'_{m+1},\hat\bp'_{m+1}) 
\mathbbm{1}_A \left(\prod_{r=1}^{m+1} 
B(\o'_r; \hp'_r-\p'_{j'_r}(t'_r))\right)\r_{n+m+1}^0(\bze'(0))\;.\nn\\
\label{eq:valtrtr}
\eea

For the well--posedness of collision histories and of the integrals in which they are involved, we need the following result.
\begin{lem} \label{lem:mp}
For any given $\TT_{n,m}$ and any $t>0$, if $\bz_n$ varies in 
$\G_n^{\dagger}$ and $(\bt_m,\hat\bp_m,\bo_m)$ in $\CC_{\TT_{n,m}}(\bz_n,t)$, the transformation 
$(\bz_n,\bt_m,\hat\bp_m,\bo_m) \longrightarrow \bze(0)$ described in Section \ref{subsec:fict}
defines $d\m_m-$almost surely a map into~$\G_{n+m}^{\dagger}$. The integrand in \eqref{eq:valtr} is a Borel function of 
$(\bz_n,\bt_m,\hat\bp_m,\bo_m)$ in the same domain.
\end{lem}
For $m=0$ this is a trivial consequence of Proposition \ref{prop:dyn}, since the transformation reduces to $T^{(n)}_{-t}$.
Therefore, we may assume the statement to be valid for $n+1$ and $m \leq N-n-1$, 
and prove it for $n$ and $m\leq N-n$, together with Proposition \ref{lem:int}. Actually, at each step of the induction,
we only need to prove the lemma for $m=N-n$. In fact, the proof of validity for given $\ol m,\ol n$,
can be applied lexicographically to $\ol m,n$ with $n$ arbitrary (just change the value of $N$).

Notice that Lemma \ref{lem:mp} (for $m=1$) implies immediately
\begin{cor} \label{cor:mp}
If $\bz_n\in\G_n^{\dagger}$, then $(T^{(n)}_{s}\bz_n, q_j^{(n)}(s)+a\o,\hat p) \in \G_{n+1}^{\dagger}$ 
for all $j=1,\cdots,n$ and almost all $(s,\o,\hat p)\in~\RRR\times\O_{j}(T^{(n)}_{s}\bz_n)\times\RRR^3$. 
\end{cor}

The proof of Proposition \ref{lem:int} is made of two steps which we separate in the two lemmas that follow. We will first prove,
in the next subsection,
\begin{lem} \label{lem:int1}
Under the assumptions of Proposition \ref{lem:int}, there holds
\be
\II(\TT)= \d_{\ell,m+1}\left( N-n -m \right) V(\TT'') +
\sum_{k = 1}^{\ell}\sum_{i = 1}^{n+k-1}\left(V |_{\FF^+_{k,i}}(\TT')+V |_{\FF^-_{k,i}}(\TT')\right)\;,
\label{eq:lemint1}
\ee
where the r.h.s. is a Borel function over $\G_n$ with absolute value bounded by $A'\prod_{j=1}^n h_{\b'}(p_j)$, for some
$A',\b'>0$.
\end{lem}

The second step will consist in showing that the collision histories which have been eliminated by the cutoff in Eq. \eqref{eq:lemint1}
give a net contribution equal to zero, i.e.
\begin{lem} \label{lem:int2}
Under the assumptions of Proposition \ref{lem:int}, there holds
\be
\sum_{k = 1}^{\ell}\sum_{i = 1}^{n+k-1}\left(V |_{\RR^+_{k,i}}(\TT')+V |_{\RR^-_{k,i}}(\TT')\right)=0\;.
\label{eq:lemint2}
\ee
\end{lem}
\noindent This will be done in a subsequent subsection.

\subsubsection*{Proof of Lemma \ref{lem:int1}}
Our task is to integrate out the variable $z_{n+1}$ in the expression (see \eqref{eq:rnesIT}, \eqref{eq:valtr})
\be
\int_{\CC_{\TT}(\bz_{n+1},t)} d\m(\bt_m,\bo_m,\hat\bp_m) \left(\prod_{r=1}^m 
B(\o_r; \hp_r-\p_{j_r}(t_r))\right)\r_{n+1+m}^0(\bze(0))\;, 
\ee
where the collision history $\bze$ is the one associated to the tree $\TT$. Since the claim in Lemma \ref{lem:mp}
is true for the considered tree and since $\r_{n+1+m}^0\in{\mathcal L}_{n+1+m}$ (remember estimate 
\eqref{eq:estwce}), the function $(z_{n+1},\bt_m,\hat\bp_m,\bo_m) \longrightarrow (\prod_rB_r)\r_{n+1+m}^0(\bze(0))$ is absolutely 
integrable over the space
\be
\CC^+ = \{z_{n+1}\in\G_1(\bz_n), (\bt_m,\hat\bp_m,\bo_m)\in\CC_{\TT}(\bz_{n+1},t)\}\;.
\ee
By Fubini's theorem, we may rewrite $\II(\TT)$  as the $6(m+1)-$dimensional integral 
\be
\II(\TT)=\int_{\CC^+} dz_{n+1}d\m_m\left(\prod_{r=1}^m B_r\right)\r_{n+1+m}^0(\bze(0))\;.
\ee

Almost surely over $\CC^+$ we have the following partition:
\be
1 = \mathbbm{1}_{\FF_0^+}+\sum_{k = 1}^{\ell}\sum_{i = 1}^{n+k-1}\mathbbm{1}_{\tilde\FF^+_{k,i}}\;,
\ee
where
\bea
&& \FF_0^+ = \Big\{ (z_{n+1},\bt_m,\hat\bp_m,\bo_m)\in\CC^+ \mbox{\ \ s.t. } 
\dist\left(\bze(s),T^{(1)}_{-t+s}z_{n+1}\right) > a\ \forall s\in(t_{\ell},t) \Big\}\;,\nn\\
&& \tilde\FF^+_{k,i} = \Big\{ (z_{n+1},\bt_m,\hat\bp_m,\bo_m)\in\CC^+ \mbox{\ \ s.t. $\exists s_+\in(t_{k},t_{k-1})$\;,\ \ } 
\dist\left(\bze(s),T^{(1)}_{-t+s}z_{n+1}\right) > a\ \forall s\in(s_+,t)\;,\nn\\
&&\ \ \ \ \ \ \ \ \ \ \ \Big|q^{(1)}_{n+1}(s_+)-\xi_i(s_+)\Big| = a\;,\ 
\left(q^{(1)}_{n+1}(s_+)-\xi_i(s_+)\right)\cdot\left(p^{(1)}_{n+1}(s_+)-\p_i(s_+)\right)>0 \Big\}\;.
\eea
The set $\FF_0^+$ collects the histories $\bze$ such that particle $n+1$ flows backwards freely up to the time in which it creates another 
particle, or up to time zero if this never happens. The set $\tilde\FF^+_{k,i}$ collects the histories such that particle $n+1$ flows 
backwards freely only up to a collision with the (pre--existent) particle $i$ of the history, occurring in the time interval $(t_{k},t_{k-1})$.
The instant of this interaction is called $s_+$.

Consider the integral
\be
\int_{\CC^+} dz_{n+1}d\m_m\mathbbm{1}_{\tilde\FF^+_{k,i}} \prod_{r=1}^m B_r \r_{n+1+m}^0(\bze(0))\;. \label{eq:intC+}
\ee
Using only the free flow of particle $n+1$, we define the change of variables
\be
(z_{n+1},\bt_m,\hat\bp_m,\bo_m) \longrightarrow (s_+,\o_+,p_+,\bt_m,\hat\bp_m,\bo_m)
\label{eq:cv1}
\ee
where $s_+$ is the time appearing in the definition of $ \tilde\FF^+_{k,i}$, and $\o_+,p_+$ describe the collision configuration
at time $s_+$, that is
\be
\o_+ = (q^{(1)}_{n+1}(s_+)-\xi_i(s_+))/a\;,\ \ \ \ \ p_+ = p^{(1)}_{n+1}(s_+)\;.
\ee
Introducing $j'_1,\cdots,j'_{m+1}$ as defined in \eqref{eq:TTp} and renaming
\bea
&& t'_1=t_1,\cdots,t'_{k-1}=t_{k-1},t'_k = s_+, t'_{k+1} = t_k,\cdots,t'_{m+1} = t_m,\nn\\
&& \o'_1=\o_1,\cdots,\o'_{k-1}=\o_{k-1},\o'_k = \o_+, \o'_{k+1} = \o_k,\cdots,\o'_{m+1} = \o_m,\nn\\
&& \hat p'_1=\hat p_1,\cdots,\hat p'_{k-1}=\hat p_{k-1},\hat p'_k = \hat p_+, \hat p'_{k+1} = \hat p_k,\cdots,\hat p'_{m+1} = \hat p_m,
\eea
we see that \eqref{eq:cv1} is an invertible transformation from $\tilde\FF^+_{k,i}$ onto $\FF^+_{k,i}$ (modulo exclusion
of sets of measure zero), i.e. the change of variables is (partially) ``generating'' the node $k$ of the tree $\TT'$. Of course, the 
transformation introduced is a Borel map. Moreover, a simple computation shows that it has Jacobian determinant given by
\be
dz_{n+1} = B(\o'_k; \hat p'_k-\p'_{j'_k}(t'_k)) dt'_kd\o'_kd\hat p'_k\;.
\ee
This $B$ factor is added to the product in \eqref{eq:intC+}, and reconstructs the factor associated to node $k$ in the formula for
$V(\TT')$. Finally, notice that the collision histories $\bze$ (associated to $\TT$) and $\bze'$ (associated to $\TT'$ and given by
the variables \eqref{eq:chprime} defined above) {\em coincide} in the time interval $(0,t'_k)$, thanks to our initial restriction to
$\tilde\FF^+_{k,i}$. Summarising, we have found
\be
\int_{\CC^+} dz_{n+1}d\m_m\mathbbm{1}_{\tilde\FF^+_{k,i}} \prod_{r=1}^m B_r \r_{n+1+m}^0(\bze(0))
= V |_{\FF^+_{k,i}}(\TT')\;.
\ee
As the above change of variables can be done for any $\bz_n\in\G_n^{\dagger}$, the transformation
$(\bz_n,\bt'_{m+1},\hat\bp'_{m+1},\bo'_{m+1})\rightarrow\bze(0)$, $\bz_n\in\G_n^{\dagger}$, 
$(\bt'_{m+1},\hat\bp'_{m+1},\bo'_{m+1})$ a.e. in $\CC_{\TT'}(\bz_n,t)\cap\FF^+_{k,i}$, associated to the tree 
$\TT'$, is into $\G_{n+m+1}^{\dagger}$ and measurable. Therefore Lemma \ref{lem:mp} is true for all trees
of the type $\TT'$ when the variables are restricted to $\FF^+_{k,i}$. Proceeding as in \eqref{eq:estwce}, we 
conclude that the integral in $V |_{\FF^+_{k,i}}(\TT')$ is absolutely convergent, and defines a Borel function 
over $\G_n$ satisfying the estimate
\be
\Big|V |_{\FF^+_{k,i}}(\TT')\Big| \leq A'\prod_{j=1}^n h_{\b'}(p_j)\;,\label{eq:estVr}
\ee
for suitable $A',\b'>0$.

Consider now the restriction to $\FF_0^+$. Here particle $n+1$ flows freely up to $t_\ell$, so that the first $n+\ell$
particles of the collision history are, at that time, 
\be
\z_1(t_\ell),\cdots,\z_n(t_\ell),T^{(1)}_{-t+t_\ell}z_{n+1},\z_{n+2}(t_\ell),\cdots,\z_{n+\ell}(t_\ell)\;.
\ee
Hence, excluding particle $n+1$ and up to a renaming of variables, the collision history in the time interval $(t_{\ell},t)$ is
equally well described by $\TT''$, see Eq. \eqref{eq:TTs}.

It is convenient to use first the measure--preserving change of variables
\be
z_{n+1} \longrightarrow \ol z= (\ol q,\ol p) = T^{(1)}_{-t+t_\ell}z_{n+1}\;.
\ee
Furthermore, unlike in the case of $ \tilde\FF^+_{k,i}$ discussed above, we shall fix the order of integration and rewrite
consequently the domains. To describe the collision history in $(t_{\ell},t)$ we will use the set of variables associated to
$\TT''$,
\bea
&& \CC_{\TT''}(\bz_{n},(t_{\ell},t)) = \Big\{(\bt''_{\ell-1},\bo''_{\ell-1},\hat\bp''_{\ell-1})\in\RRR^{\ell-1}\times 
S^{2({\ell-1})}\times\RRR^{3({\ell-1})}\ \Big|\ \nn\\
&& \ \ \ \ \ \ \ \ \ \ \ \ \ \ \ \ \ \ \ \ \ \ \ \ \ \ t=t''_0>t''_1>\cdots>t''_{\ell-1}>t''_{\ell}=t_{\ell},\ \o''_k\in\O_{j''_k}(\bze''_{n+k-1}(t''_k))\Big\}\;,
\eea
while, to describe the history in the time interval $(0,t_{\ell})$, we will use the set of variables associated to the auxiliary tree 
$\TT''' =~\TT'''_{n+\ell+1,m-\ell} =~\bj'''_{m-\ell}$,
\bea
&& \bj'''_{m-\ell} = f'''(j_{\ell+1}),\cdots,f'''(j_m) \nn\\
&& f'''(j) = 
\left\{
\begin{array}{cc}
f''(j)  &  \mbox{ \ \ \ \ \ \ \ \ \ \ \ \ \ \ \ \ \ \ \ \ \ if \ \ } j\leq n+\ell,\ j\neq n+1   \\
n+\ell & \mbox{ \ \ \ \ \ \ \ if \ \ } j = n+1    \\
j & \mbox{ \ \ \ \ \ \ \ \ \ \ \ \ if \ \ } j\geq n+\ell +1
\end{array}
\right.\;.
\eea
With these notations we have
\bea
&& \int_{\CC^+} dz_{n+1}d\m_m\mathbbm{1}_{\FF^+_0} \left(\prod_{r=1}^m B_r\right) \r_{n+1+m}^0(\bze(0))
= \int_{\CC_{\TT''}(\bz_{n},(t_{\ell},t))}d\m''_{\ell-1} \left(\prod_{r=1}^{\ell-1} B''_r \right)\int_0^{t''_{\ell-1}} dt_{\ell}\nn\\
&& \ \ \ \ \ \ \ \ \ \ \ \ \ \ \ \ \ \ \ \ \ \ \ \ \ \ \ \ \ \ \ \ \ \ \ \ \ \ \ \ \ \ \ \cdot \int_{\G_1(\bze''_{n+\ell-1}(t_{\ell}))} d\ol z
\ \mathbbm{1}_{\{\dist(\bze''(s),\ol z^{(1)}(s))>a\ \forall s\in(t_{\ell},t)\}}\nn\\
&&  \ \ \ \ \ \ \ \ \ \ \ \ \ \ \ \ \ \ \ \ \ \ \ \ \ \ \ \ \ \ \ \ \ \ \ \ \ \ \ \ \ \ \ 
\cdot\int_{\{S^2\times\RRR^3,\ \bz'''_{n+\ell+1} \in \G_{n+\ell+1}\}} d\o_{\ell}d\hat p_{\ell}
\ B(\o_{\ell}; \hat p_{\ell}-\ol p)\nn\\
&& \ \ \ \ \ \ \ \ \ \ \ \ \ \ \ \ \ \ \ \ \ \ \ \ \ \ \ \ \ \ \ \ \ \ \ \ \ \ \ \ \ \ \ 
\cdot \int_{\CC_{\TT'''}(\bz'''_{n+\ell+1},t_{\ell})} d\m'''_{m-\ell} \left(\prod_{r=1}^{m-\ell} B'''_r\right)
\r_{n+m+1}^0(\bze'''(0))\;, \label{eq:F0+f}
\eea
where $\ol z^{(1)}(s)=T^{(1)}_{-t_{\ell}+s}\ol z$,
\be
\bz'''_{n+\ell+1}=\bze''_{n+\ell-1}(t_{\ell}),\ol z,\ol q+a\o_{\ell},\hat p_{\ell}\;,
\ee
$\bze''$ is the history associated to $\TT''$, $d\m''_{\ell-1} = d\m(\bt''_{\ell-1},\bo''_{\ell-1},\hat\bp''_{\ell-1})$, 
$B''_r = B(\o''_r; \hat p''_r-\p''_{j''_r}(t''_r))$, etc. The last line of \eqref{eq:F0+f} is just $V(\TT''')$
evaluated in the collision configuration $\bz'''_{n+\ell+1}$ at time $t_{\ell}$. Of course, in the case
$\ell = m+1$, the expression is much simpler, since $t_{\ell}\equiv 0$ and there are no variables
$\o_{\ell},\hat p_{\ell}, \cdots$ (see formula \eqref{eq:simplet} below).

Call now
\be
\CC^- = \{(\bt''_{\ell-1},\hat\bp''_{\ell-1},\bo''_{\ell-1})\in\CC_{\TT''}(\bz_{n},(t_\ell,t)),\ \ol z \in\G_1(\bze''_{n+\ell-1}(t_{\ell}))\}\;.
\ee
In formula \eqref{eq:F0+f} we can write
\be
\mathbbm{1}_{\{\dist(\bze''(s),\ol z^{(1)}(s))>a\ \forall s\in(t_{\ell},t)\}} = 1 -\sum_{k=1}^{\ell}
\sum_{i = 1}^{n+k-1}\mathbbm{1}_{\tilde\FF^-_{k,i}} \label{eq:secdec}
\ee
with
\bea
&& \tilde\FF^-_{k,i} = \Big\{ (\ol z,\bt''_{\ell-1},\hat\bp''_{\ell-1},\bo''_{\ell-1})\in\CC^- \mbox{\ \ s.t. $\exists s_-\in(t''_{k},t''_{k-1})$\;,\ \ } 
\dist\left(\bze''(s),\ol z^{(1)}(s)\right) > a\ \forall s\in(t_\ell,s_-)\;,\nn\\
&& \ \ \ \ \ \ \ \ \ \ \ \Big|\ol q^{(1)}(s_-)-\xi''_i(s_-)\Big| = a\;,\ 
\left(\ol q^{(1)}(s_-)-\xi''_i(s_-)\right)\cdot\left(\ol p^{(1)}(s_-)-\p''_i(s_-)\right)<0 \Big\}\;.
\eea
That is, we add and subtract the sets of variables such that a particle with state $\ol z$ collides, when evolved
freely forward in time, with one of the particles of $\bze''$. We name $s_-$ the instant of this interaction.

We shall see that in the (added and) subtracted restrictions to $\tilde\FF^-_{k,i}$, the collision histories are well--defined. 
In fact, since $\TT''$ has less than $m\leq N-n-1$ nodes, $\bze''_{n+\ell-1}(t_{\ell})\in\G_{n+\ell-1}^{\dagger}$
almost surely with respect to $d\m''_{\ell-1}$ (apply Lemma \ref{lem:mp}). Of course an analogous property holds for $\bze''$
at different times. Thus, for any given $t_{\ell}$ and $d\m''_{\ell-1}-$a.e., there holds $(\bze''_{n+\ell-1}(t_{\ell}),\ol z)\in\G_{n+\ell}^\dagger$ 
for $\ol z$ in a full measure subset of $\G_1(\bze''_{n+\ell-1}(t_{\ell}))$. If $n=N-1$, this is enough ($m=0, \ell=1$, no histories of
type $\bze'''$). Otherwise, to deal with the case $\ell< m+1$, we apply Corollary \ref{cor:mp} to any configuration
$(\bze''_{n+\ell-1}(t_{\ell}),\ol z)\in\G_{n+\ell}^\dagger$. This implies that $\bz'''_{n+\ell+1}\in\G_{n+\ell+1}^\dagger$
almost everywhere with respect to the measure $d\m''_{\ell-1}dt_\ell d\ol z d\o_{\ell}d\hat p_{\ell}$. Using Lemma \ref{lem:mp}
for the tree $\TT'''$, we deduce that $\bze'''(0)\in\G_{n+m+1}^\dagger$ almost everywhere in the domain of integration
(and analogous property for $\bze'''$ at different times), the last line of \eqref{eq:F0+f} is well--defined and all the integrals are 
absolutely convergent.

Each restriction to $\tilde\FF^-_{k,i}$ can be treated as we did for $\tilde\FF^+_{k,i},$ i.e. with a change of variables
\be
\ol z \longrightarrow (s_-,\o_-,p_-) 
\ee
where
\be
\o_- = (\ol q^{(1)}(s_-)-\xi''_i(s_-))/a\;,\ \ \ \ \ p_- = \ol p^{(1)}(s_-)\;.
\ee
Introducing $\bj'_{m+1}$ as defined in \eqref{eq:TTp}, renaming
\bea
&& \bt'_{k-1}=\bt''_{k-1}, t'_k = s_-, t'_{k+1} = t''_k,\cdots,
t'_\ell=t''_{\ell-1},t'_{\ell+1}=t_\ell,t'_{\ell+2}=t'''_1,\cdots,t'_{m+1} = t'''_{m-\ell},\nn\\
&& \bo'_{k-1}=\bo''_{k-1}, \o'_k = \o_-, \o'_{k+1} = \o''_k,\cdots,
\o'_\ell=\o''_{\ell-1},\o'_{\ell+1}=\o_\ell,\o'_{\ell+2}=\o'''_1,\cdots,\o'_{m+1} = \o'''_{m-\ell},\nn\\
&& \hat\bp'_{k-1}=\hat\bp''_{k-1}, \hat p'_k = p_-, \hat p'_{k+1} = \hat p''_k,\cdots,
\hat p'_\ell=\hat p''_{\ell-1},\hat p'_{\ell+1}=\hat p_\ell,\hat p'_{\ell+2}=\hat p'''_1,\cdots,\hat p'_{m+1} = \hat p'''_{m-\ell},
\eea
and excluding sets of measure zero, we see that the above change of variables defines an invertible transformation 
from $\tilde\FF^-_{k,i}$ onto $\FF^-_{k,i}$, with
\be
d\ol z = - B(\o'_k; \hat p'_k-\p'_{j'_k}(t'_k)) dt'_kd\o'_kd\hat p'_k\;.
\ee
Summarising, Lemma \ref{lem:mp} is true for all trees of the type $\TT'$ when the variables are restricted to 
$\FF^-_{k,i}$, and 
\be
- \int_{\CC^+} dz_{n+1}d\m_m\mathbbm{1}_{\tilde\FF^-_{k,i}} \left(\prod_{r=1}^m B_r\right) \r_{n+1+m}^0(\bze(0))
= V |_{\FF^-_{k,i}}(\TT')\;,
\ee
with this satisfying the same estimate of $V |_{\FF^+_{k,i}}$ in \eqref{eq:estVr}.

So far we have proved Lemma \ref{lem:mp} for all trees $\TT'$ with a restriction of the $k-$th node variables given by the 
definitions of $\FF^+_{k,i}, \FF^-_{k,i}$. Observe that this restriction can be immediately eliminated by using the arbitrariness
of the time interval $(0,t)$ and the invariance of the set $\G_{n+m}^{\dagger}$. 
Varying $\TT_{n+1,m}$ in the hypotheses of Proposition \ref{lem:int}, we conclude that Lemma 
\ref{lem:mp} holds for all $\TT_{n,m}$ with $m\leq N-n$.

To prove Lemma \ref{lem:int1}, we are left with the term ``$1$'' in \eqref{eq:secdec}. There are two cases.

{\em Case $\ell = m+1$.} Formula $\eqref{eq:F0+f}$, without the cutoff $\mathbbm{1}$, reduces to 
\be
\int_{\CC_{\TT''}(\bz_{n},t)}d\m''_m \left(\prod_{r=1}^m B''_r \right) 
\int_{\G_1(\bze''_{n+m}(0))} d\ol z\  \r_{n+m+1}^0(\bze''(0),\ol z)\;. \label{eq:simplet}
\ee
Using Eq. \eqref{eq:defcralt}, this gives the term $\left( N-n -m \right) V(\TT'')$.

{\em Case $\ell < m+1$.} Formula $\eqref{eq:F0+f}$, without the cutoff $\mathbbm{1}$, reduces to zero. Indeed, consider
\be
\int_{\{\RRR^6\times S^2\times\RRR^3,\ \bz'''_{n+\ell+1}\in\G_{n+\ell+1}\}} d\ol z d\o_{\ell}d\hat p_{\ell}
\ B(\o_{\ell}; \hat p_{\ell}-\ol p)\ V(\TT''')(\bz'''_{n+\ell+1}, t_{\ell})\;. \label{eq:zeropartes}
\ee
Almost surely over the domain, the elastic scattering defines a one--to--one mapping between outgoing and incoming
collision configurations $(\ol z,\ol q+a\o_\ell,\ol p)$. Under this mapping the factor $B$ in \eqref{eq:zeropartes} changes
sign, while $V(\TT''')$ is preserved.

Summing all contributions, we obtain Eq. \eqref{eq:lemint1}. This ends the proof of Lemma \ref{lem:int1}.
$\hfill\Box$

\subsubsection*{Proof of Lemma \ref{lem:int2}}

In what follows we shall indicate explicitly as $\TT'(k,i)$ the dependence on $k,i$ of the trees of type $\TT'$.

Let us focus on $\TT'(k,i),\RR^-_{k,i}$. This collects the collision histories such that particle $n+k$, after having been generated
by particle $i$ in an incoming collision, ``recollides'' with some other particle of the history (see the comment before
\eqref{eq:expift} about this terminology). 
Given one of such histories, let us {\em erase} the {\em free} flow of particle $n+k$ from the moment of 
generation ($t'_k$) to the moment of recollision, and think that the particle {\em appears} at the recollision time in an outgoing collision 
configuration. In other words, we transform the recollision in a creation. What we obtain is a new collision history, which will
be associated to some $\TT'(k^*,i^*)$ and will obey the constraint of $\RR^+_{k^*,i*}$. Roughly speaking, the two related collision 
histories ``cancel'' each other in the computation of the left hand side of \eqref{eq:lemint2}.

To make precise the last assertion, we decompose further the domains by specifying which particle recollides with $n+k$ and in which
time interval the recollision occurs. Fixed a tree $\TT'(k,i)$, we introduce a set $\RR^-_{k,i;k^*,j}$ selecting the collision histories
such that (i) particle $n+k$ is generated by particle $i$ in an incoming collision; (ii) particle $n+k$ recollides with particle $j$ of the
history; (iii) such a recollision occurs in the time interval of the history $(t'_{k^*+1},t'_{k^*})$. A similar notation is introduced for
the $+$ case. In formulas,
\bea
&& \sum_{k = 1}^{\ell}\sum_{i = 1}^{n+k-1}\left(V |_{\RR^+_{k,i}}(\TT'(k,i))+V |_{\RR^-_{k,i}}(\TT'(k,i))\right)\nn\\
&& =   \sum_{1 \leq k\leq k^*\leq\ell} \sum_{i = 1}^{n+k-1}\Big[\Big(\sum_{j = 1}^{n+k^*-1}V |_{\RR^+_{k^*,j;k,i}}(\TT'(k^*,j))\Big)+
\Big(\underset{j\neq n+k}{\underset{j=1}{\sum^{n+k^*}}}V |_{\RR^-_{k,i;k^*,j}}(\TT'(k,i))\Big)\Big]\label{eq:cdrkks}
\eea
where
\bea
&& \RR^+_{k^*,j;k,i} =  \Big\{ (\bt'_{m+1},\bo'_{m+1},\hat\bp'_{m+1})\in\RR^+_{k^*,j} \mbox{\ \ s.t.\ \ } 
s_+\in(t'_{k},t'_{k-1}),\ i_+=i \Big\}\;,\nn\\
&& \RR^-_{k,i;k^*,j} = \Big\{ (\bt'_{m+1},\bo'_{m+1},\hat\bp'_{m+1})\in\RR^-_{k,i} \mbox{\ \ s.t.\ \ } 
s_-\in(t'_{k^*+1},t'_{k^*}),\ i_-=j \Big\}\;. \label{eq:RRkiksj}
\eea
In \eqref{eq:RRkiksj} $s_+,i_+,s_-,i_-$ are those appearing in the definition of $\RR^+_{k^*,j}, \RR^-_{k,i}$.
Notice that, in the second sum over $j$ of \eqref{eq:cdrkks}, the value $n+k$ is obviously missing, since particle $n+k$ cannot
recollide with itself.

Fix an integral term $V |_{\RR^-_{k,i;k^*,j}}$ of the above sum. Remember that this is an integral over a subset of 
$\CC_{\TT'(k,i)}(\bz_{n},t)$, i.e. the node--variables associated to the tree $\TT'(k,i)$. We change the variables of integration
according to
\be
(t'_k,\o'_k,\hat p'_k) \longrightarrow (s_-,\o_-,p_-)\;,
\ee
where $s_-$ is defined in \eqref{eq:defR+R-}, \eqref{eq:RRkiksj}, and
\be
\o_- = (\xi'_{n+k}(s_-)-\xi'_j(s_-))/a\;,\ \ \ \ \ p_- = \p'_{n+k}(s_-)\;.
\ee
With the renaming 
\bea
&& \bt'_{k-1}\rightarrow\bt'_{k-1}, (t'_{k+1},\cdots,t'_{k^*})\rightarrow (t'_{k},\cdots,t'_{k^*-1}),
s_- = t'_{k^*}, (t'_{k^*+1},\cdots,t'_{m+1})\rightarrow (t'_{k^*+1},\cdots,t'_{m+1}), \nn\\
&& \bo'_{k-1}\rightarrow\bo'_{k-1}, (\o'_{k+1},\cdots,\o'_{k^*})\rightarrow (\o'_{k},\cdots,\o'_{k^*-1}),
\o_- = \o'_{k^*}, (\o'_{k^*+1},\cdots,\o'_{m+1})\rightarrow (\o'_{k^*+1},\cdots,\o'_{m+1}), \nn\\
&& \hat\bp'_{k-1}\rightarrow\hat\bp'_{k-1}, (\hat p'_{k+1},\cdots,\hat p'_{k^*})\rightarrow (\hat p'_{k},\cdots,\hat p'_{k^*-1}),
\hat p_- = \hat p'_{k^*}, (\hat p'_{k^*+1},\cdots,\hat p'_{m+1})\rightarrow (\hat p'_{k^*+1},\cdots,\hat p'_{m+1}), \nn\\
\eea
we obtain an invertible map (modulo sets of measure zero) onto $\RR^+_{k^*,i^*;k,i}$, that is
a subset of the node--variables associated to the tree $\TT'(k^*,i^*)$, with
\be
i^* = 
\left\{
\begin{array}{cc}
j  &  \mbox{ \ \ \ \ \ \ \ \ if \ \ } j < n+k   \\
j-1 & \mbox{ \ \ \ \ \ \ \ \ if \ \ }  j > n+k    \\
\end{array}
\right.\;.
\ee

Now, observe that the Jacobian determinant is given by the relation
\be
- B(\o'_k; \hat p'_k-\p'_{j'_k}(t'_k)) dt'_kd\o'_kd\hat p'_k = B(\o_-; p_- - \p'_{j}(s_-)) ds_-d\o_- dp_-\;,
\ee
the minus sign coming from the fact that the variables appearing in the l.h.s. describe an incoming collision, while
the variables appearing in the r.h.s. describe an outgoing collision. Therefore, the net effect of the transformation is
\be
V |_{\RR^-_{k,i;k^*,j}}(\TT'(k,i)) = - V |_{\RR^+_{k^*,i^*;k,i}}(\TT'(k^*,i^*))\;.
\ee
Inserting this into Eq. \eqref{eq:cdrkks}, we obtain Lemma \ref{lem:int2}.
$\hfill\Box$

\subsection{The sum over trees}\label{ss:sum}

To prove Theorem \ref{thm:result}, it remains to substitute Eq. \eqref{eq:ITrule} into Eq. \eqref{eq:sIIT} and perform the sum
over trees. This can be achieved conveniently by working directly on the graphs, as shown in Figure \ref{fig:proof}. The rules given
by the list on page \pageref{list:rules} tell us which trees appear on the right hand side of \eqref{eq:ITrule}. (Notice that,
by applying the rule in step 1 with different values of $k,i$, a tree $\TT_{n+1,m}$ can even produce more copies of the same 
tree $\TT'_{n,m+1}$.) Hence, it is sufficient to check that any given $n-$particle, $m-$node tree $\TT_{n,m}$ can be produced in 
exactly $N-n$ copies, by applying the rules to different $(n+1)-$particle trees. This follows from the remarks: (i) $\TT_{n,m}$
is produced in $N-n-m$ copies by operation 3 of the list; (ii)~$\TT_{n,m}$ is produced by creating its node $k$, applying 
operation 1 of the list to a suitable $(n+1)-$particle tree. Summing up, we have $N-n-m+m=N-n$ copies.

The analogous algebraic proof is as follows:
\bea
&& \r_n(\bz_n,t) =  \frac{1}{N-n}\sum_{m=0}^{\infty} \sum_{\TT_{n+1,m}} \Big[\d_{\ell,m+1}\left( N-n -m \right) V(\TT''_{n,m}) +
\sum_{k = 1}^{\ell}\sum_{i = 1}^{n+k-1}V(\TT'_{n,m+1})\Big]\nn\\
&& = \frac{1}{N-n}\sum_{m=0}^{\infty}(N-n -m) \sum_{\substack{j_1,\cdots,j_m\\ j_r\in I_{n+r}\\ j_r\neq n+1}}V(\TT''_{n,m})
+  \frac{1}{N-n}\sum_{m=1}^{\infty}\sum_{\substack{j_1,\cdots,j_{m-1}\\ j_r\in I_{n+r}}}\sum_{k = 1}^{\ell}\sum_{i = 1}^{n+k-1}V(\TT'_{n,m})\nn\\
&& = \frac{1}{N-n}\sum_{m=0}^{\infty}(N-n -m) \sum_{\substack{j''_1,\cdots,j''_m\\ j''_r\in I_{n+r-1}}}V(\TT''_{n,m})
+ \frac{1}{N-n}\sum_{m=1}^{\infty}\sum_{k = 1}^{m}\sum_{\substack{j_1,\cdots,j_{m-1}\\ j_r\in I_{n+r}\\j_1,\cdots,j_{k-1}\neq n+1}}
\sum_{i = 1}^{n+k-1}V(\TT'_{n,m})\nn\\
&& = \frac{1}{N-n}\sum_{m=0}^{\infty}(N-n -m) \sum_{\TT''_{n,m}}V(\TT''_{n,m})
+ \frac{1}{N-n}\sum_{m=1}^{\infty}\sum_{k = 1}^{m}\sum_{\substack{j'_1,\cdots,j'_{m}\\ j'_r\in I_{n+r-1}}}V(\TT'_{n,m})\nn\\
&& = \frac{1}{N-n}\sum_{m=0}^{\infty}(N-n -m) \sum_{\TT''_{n,m}}V(\TT''_{n,m})
+ \frac{1}{N-n}\sum_{m=1}^{\infty}m\sum_{\TT'_{n,m}}V(\TT'_{n,m})\nn\\
&& = \sum_{m=0}^{\infty}\sum_{\TT_{n,m}}V(\TT_{n,m})\;,
\eea
where in the third and in the fourth line we have used respectively the definitions \eqref{eq:TTs} and \eqref{eq:TTp}.
$\hfill\Box$

\subsection{The BBGKY hierarchy. Proof of Corollary \ref{cor:resultB}} \label{sec:resultB}

Let us rewrite the expansion \eqref{eq:expift} in a resummed form, which is convenient to obtain informations about 
the derivative. 

We have proven that the integrals \eqref{eq:valtr} are absolutely convergent, so that the integration
order can be exchanged freely. Then, in the hypotheses of Corollary \ref{cor:resultB}, fixed $\bz_n\in\G_n^{\dagger}$ 
and $t>0$, we have
\bea
&& \r_n(\bz_n,t) = \TT_{n,0}(\bz_n,t) + \sum_{\TT_{n,1}} V(\TT_{n,1})(\bz_n,t) + \sum_{m>1} \sum_{\TT_{n,m}} V(\TT_{n,m})(\bz_n,t)\nn\\
&&\ \ \ \ \ \ \ \ \ \ \ \ = \r_n^0(T^{(n)}_{-t}\bz_n) + \sum_{j_1=1}^n \int_{(0,t)\times\RRR^3\times\O_{j_1}(T^{(n)}_{-t+t_1}\bz_n)} 
dt_1 d\hat p_1 d\o_1 \ a^2 \o_1\cdot \left(\hat p_1 - p_{j_1}^{(n)}(t_1)\right) \nn\\
&&\ \ \ \ \ \ \ \ \ \ \ \ \ \ \ \ \ \ \ \ \ \ \ \ \ \ \ \ \ \ \ \cdot \Bigg[\r_{n+1}^0\left(T^{(n+1)}_{-t_1} \left(T^{(n)}_{-t+t_1}\bz_n, q_{j_1}^{(n)}(t_1)
+a\o_1,\hat p_1\right)\right) \nn\\
&&\ \ \ \ \ \ \ \ \ \ \ \ \ \ \ \ \ \ \ \ \ \ \ \ \ \ \ \ \ \ \ + \sum_{m \geq 1}\sum_{\TT_{n+1,m}}V\left(\TT_{n+1,m}\right)
\left(T^{(n)}_{-t+t_1}\bz_n, q_{j_1}^{(n)}(t_1)+a\o_1,\hat p_1,t_1\right) \Bigg]\;,
\eea
where $q_{j_1}^{(n)}(t_1),p_{j_1}^{(n)}(t_1)$ are position and momentum of particle $j_1$ in $T^{(n)}_{-t+t_1}\bz_n$.
In the last term we have put together the one--node trees and the higher order trees, $dt_1 d\hat p_1 d\o_1$ being the 
integration associated to the first node.

By Corollary \ref{cor:mp}, we may use again Equation \eqref{eq:expift} to identify the term in the square brackets 
with a $\r_{n+1}(\cdot,t_1)$, that is
\bea
&& \r_n(\bz_n,t) = \r_n^0(T^{(n)}_{-t}\bz_n) + \sum_{j_1=1}^n \int_{(0,t)\times\RRR^3\times\O_{j_1}(T^{(n)}_{-t+t_1}\bz_n)} 
dt_1 d\hat p_1 d\o_1 \ a^2 \o_1\cdot \left(\hat p_1 - p_{j_1}^{(n)}(t_1)\right) \nn\\
&& \ \ \ \ \ \ \ \ \ \ \ \ \ \ \ \ \ \ \ \ \ \ \ \ \ \ \ \ \ \ \ \cdot \r_{n+1}\left(T^{(n)}_{-t+t_1}\bz_n, q_{j_1}^{(n)}(t_1) + a\o_1,\hat p_1, t_1\right)\;.
\label{eq:resform}
\eea
This formula is the {\em resummed} form of the expansion for the correlation functions, in the sense that iterating the equation 
$N-n$ times we are back to the Equation \eqref{eq:expift}.

Remind now that the Liouville equation can be also written as $f_N(T_t^{(N)}\bz_N,t) = f_N(\bz_N)$ and that, being 
$\G_n^{\dagger}$ invariant, $\r_n(T_t^{(n)}\bz_n,t) = N \dots (N-n+1) \int_{\G_{N-n}(T_t^{(n)}\bz_n)}dz_{n+1}\dots dz_N 
f_N(T_t^{(n)}\bz_n,z_{n+1},\cdots,z_N,t)$ for all $\bz_n\in\G_n^{\dagger}$. In particular, we may substitute 
$\bz_n\rightarrow T_t^{(n)}\bz_n$ in \eqref{eq:resform}. Recalling \eqref{eq:Qdef}, we obtain
\bea
\r_n(T^{(n)}_{t}\bz_n,t) = \r_n(\bz_n,0) + \int_0^t dt_1 \left(Q\r_{n+1}\right) (T^{(n)}_{t_1}(\bz_n),t_1) \label{eq:BBGKYint}
\eea
where, by Fubini's theorem, the integral in $dt_1$ is well defined. Eq. \eqref{eq:BBGKYint} shows that, for all 
$\bz_n\in\G_n^{\dagger}$, the function $t \rightarrow (Q\r_{n+1}) (T^{(n)}_{t}(\bz_n),t)$ is absolutely continous, with
derivative satisfying \eqref{eq:BBGKY} for almost all times.$\hfill\Box$

\subsection{Indefinite number of particles. Proof of Corollary \ref{cor:result}} \label{sec:resultN}

Each term in the sum in Equation \eqref{eq:defcftris} may be dealed with the procedure explained in the previous sections.
This leads directly to a tree expansion like the one in the right hand side of \eqref{eq:expift}, in which the value of the tree, say 
$\tilde V(\TT_{n,m})$, must be computed in a slightly different way. Namely, $\r_{n+m}^0$ in \eqref{eq:valtr} is replaced by 
\be
\frac{1}{(k-m)!} \int_{\G_{k-m}(\bze(0))} dz_{n+m+1}\cdots dz_{n+k} f_{n+k}^0(\bze(0), z_{n+m+1},\cdots , z_{n+k})\;.
\ee
Performing the sum over $k$, that is $\sum_{k \geq m}^{\infty}$, and using \eqref{eq:defcftris}, we recover Eq. \eqref{eq:expift}.
$\hfill\Box$

\section{Conclusions} \label{sec:conc}

In this work we discussed a derivation of the series expansion 
used by Lanford \cite{Lanford} to perform the Boltzmann--Grad limit, expressing the time--evolved $n-$point
correlation function in terms of the higher order correlation functions at time zero for a system of $N$
hard spheres in a finite volume. We established a method of construction of the series based on step by step
direct integration of degrees of freedom from the solution of Liouville equation, rather than the usual iteration of
the BBGKY equations. 
Each term of the expansion was written in the form of integral over a class of special evolutions of particles
called ``collision histories'', for which we could introduce a convenient graphical representation.
These graphs are useful to control the integration procedure leading from the expansion for
$\r_{n+1}$ to the expansion for $\r_n$. Mutual cancellations between collision histories showing special
``recollision properties'' were exhibited as an important part of the proof. 

The method provides a construction of the series expansion in a fixed full measure subset of the phase space,
under the only hypotheses of some integrability bound for the density of the initial measure,
and symmetry in the particle labels. This strengthens results previously obtained in literature. Furthermore,
without assuming continuity along trajectories of the initial measure, we could resum the final expansion and 
recover the BBGKY hierarchy of integro--differential equations for hard spheres. 
Finally, we stated an extension of the results to initial measures with non definite number of particles.

\vspace{0.7cm}
{\bf Acknowledgements.} The author thanks J. L. Lebowitz for his invitation at Rutgers University, where
the idea of this work was conceived, and acknowledges G. Gallavotti for proposing the work and for 
fundamental suggestions. The author thanks also G. Genovese, G. Gentile, A. Giuliani, A. Pellegrinotti, 
M. Pulvirenti, C. Saffirio and, in particular, H. Spohn for helpful discussions and encouragement.


\appendix

\section*{Appendix. On the dynamics of hard spheres} \label{app:dynamics}
\renewcommand{\theequation}{\ref{app:dynamics}.\arabic{equation}}


In this appendix we prove Proposition \ref{prop:dynbis}. Unfortunately, it is not clear whether the 
two sets $\G_n^\dagger$ and $\G_n^*$ coincide. Therefore, we will deduce Proposition \ref{prop:dynbis}
from Proposition \ref{prop:dyn} by using an abstract argument. 

It is sufficient to prove the assertion for any finite bound on the energy.
A little abuse of notation will be used in this section: we indicate with the usual symbols $\G_n,\G_n^*,\G_n^{\dagger}  ...$ 
the bounded sets corresponding to an energy of the (whole) system not larger than $E > 0.$
We denote with $| \cdot |$ the Lebesgue measure on $\RRR^{6n}$ and with $| \cdot |_{\mbox{ext}}$ 
the associated outer measure, defined as $|A|_{\mbox{ext}} = \inf_{\{C_n\}_{n\geq 1}} \sum_n |{C}_n|$ 
where the infimum is taken over all possible collections of boxes such that $A\subset \cup_n{C}_n.$ 
The proof will make use of two simple properties of the outer measure: first, the
flow preserves outer measure, i.e.
\be
|A|_{\mbox{ext}} = |T^{(n)}_tA|_{\mbox{ext}}\;,\ \ \ \ \ \ \ \ \ \ \ A\subset\G_n^*\;,
\label{eq:presom}
\ee
which follows from the fact that the flow is an invertible and measure preserving transformation; second,
if $B_{\bz_n}$ is a collection of sets in $\RRR^{6k}$ indexed by $\bz_n\in A\subset\RRR^{6n}$ and such that 
$|B_{\bz_n}|>0$ uniformly in $A$, then
\be
|A|_{\mbox{ext}} \neq 0\hspace{0,8mm}\Longrightarrow 
\Big|\Big\{(\bz_n,\by_k)\ \Big|\ \bz_n \in A, \by_k\in B_{\bz_n}\Big\}\Big|_{\mbox{ext}} \neq 0\;.
\label{eq:abspoint}
\ee

Let us define the ``bad sets of adjoint points''
\be
B_{k,\bz_n}= \Big\{ \by_k \in \G_k(\bz_n) \ \Big|\ (\bz_n,\by_k) \in \G_{n+k}\setminus\G_{n+k}^*\Big\}\;.
\ee
As a consequence of Proposition \ref{prop:dyn}, the following subset of $\G_n$ must be null:
\be
Z = \bigcup_{k=1}^{N-n}Z_k\;,\ \ \ \ \ Z_k = \Big\{ \bz_n\in\G_n^* \ \Big|\  |B_{k,\bz_n}|_{\mbox{ext}} > 0\Big\}
\ee
(otherwise, by \eqref{eq:abspoint} we could find a subset of $\G_{n+k}\setminus\G_{n+k}^*$ of positive outer measure). 

We do not know if $Z$ is invariant under the flow. Nevertheless, to conclude the proof, it is enough to show that
\be
\Big|\bigcup_{s\in\RRR}T^{(n)}_s Z\Big|_{\mbox{ext}} = 0\;, \label{eq:dynnull}
\ee
since then the complement of this set in $\G_n^*$ would satisfy all the properties stated in the proposition.
Given any sequence of positive numbers $\e_q\rightarrow 0,$ it is thus sufficient to prove that
\be
\Big|\bigcup_{s}T^{(n)}_s Z_{k,q}\Big|_{\mbox{ext}} = 0\;,\ \ \ \ \ 
Z_{k,q} = \Big\{ \bz_n\in\G_n^* \ \Big|\  |B_{k,\bz_n}|_{\mbox{ext}} > \e_q\Big\}\;. \label{eq:TsZkq}
\ee

For $\bz_n\in\G_n^*,\ \by_k\in\G_k^*\cap\G_k(\bz_n),$ we define the time of first forward interaction between 
$\bz_n$ and $\by_k$
\be
\tau(\bz_n;\by_k) =  \inf\Big\{ t>0 \ \Big|\ T^{(n+k)}_t(\bz_n,\by_k)=
\left(T^{(n)}_t\bz_n,T^{(k)}_t\by_k\right)\Big\}\;,
\ee
and we call
\be
B_{k,\bz_n}^{(\d)}=B_{k,\bz_n}\bigcap\Big\{\by_k\in\G_k^*\cap\G_k(\bz_n) \ \Big|\ \tau(\bz_n;\by_k)>\d\Big\}\;,\ \ \ \ \ \ \ \ \ \ \d>0\;.
\ee
Observe that, by the bound on the energy, the set of values of $\by_k$ in the domain of $\tau$ such that 
$\tau(\bz_n;\by_k)\leq\d$ has a measure that goes to zero with $\d,$ uniformly in $\bz_n.$
Hence we can find a $\d_q>0$ such that $|B_{k,\bz_n}^{(\d_q)}|_{\mbox{ext}}>\e_q/2$ 
for all $\bz_n\in Z_{k,q}.$ For such a choice we deduce that
\be
0 = \Big|\bigcup_{s\in[0,\d_q]}\bigcup_{\bz_n\in Z_{k,q}}
T^{(n+k)}_s\left(\bz_n, B_{k,\bz_n}^{(\d_q)}\right)\Big|_{\mbox{ext}}
=\Big|\bigcup_{s\in[0,\d_q]}\bigcup_{\bz_n\in Z_{k,q}}
\left(T^{(n)}_s\bz_n, T^{(k)}_s B_{k,\bz_n}^{(\d_q)}\right)\Big|_{\mbox{ext}}\;, \label{eq:dynproofcr}
\ee
where the first equality is true because the set is contained in $\G_{n+k}\setminus\G_{n+k}^*$ (applying again 
Proposition \ref{prop:dyn}). By \eqref{eq:presom}, $| T^{(k)}_s B_{k,\bz_n}^{(\d_q)}|_{\mbox{ext}}>\e_q/2$. 
Therefore by \eqref{eq:abspoint} we have that 
%
\be
\Big|\bigcup_{s\in[0,\d_q]}T^{(n)}_s Z_{k,q}\Big|_{\mbox{ext}} = 0\;.
\ee
%
Since $\bigcup_{s}T^{(n)}_s Z_{k,q}=\bigcup_{j \in\ZZZ} T_{j\d_q}^{(n)}\bigcup_{s\in[0,\d_q]}T^{(n)}_s Z_{k,q},$
Eq. \eqref{eq:TsZkq} follows. The proof of Proposition \ref{prop:dynbis} is complete. 
$\hfill\Box$

\vspace{2mm}
We add now some other useful remark concerning the dynamics of hard spheres. 
Consider the set of ``collision surfaces'', i.e. the boundary of the phase space $\partial\G_n$. On it we define
the induced Lebesgue measure $d\s (\bz_n)$. The restriction of $d\s (\bz_n)$ to the set where particles 
$i$ and $j$ are colliding, with $q_j = q_i + a\o$, is $dz_1\cdots dz_i \cdots dz_{j-1} dp_j d\o dz_{j+1}\cdots dz_n$,
while the restriction to the set in which particle $i$ is colliding with the wall, $q_i = q + (a/2) n(q), q\in\partial\L$, is
$ dz_1\cdots dz_{i-1}dq dp_i dz_{i+1}\cdots dz_n$.
Of course the prescription assigns measure zero to the set of multiple collisions, grazing collisions and singular collisions 
with the wall. 
Let us call $\partial\G_n^+$ $(\partial\G_n^-)$ the subset of points that can be reached continuously from the interior 
of $\G_n$ through the backwards (forward) free flow. $\partial\G_n^+$ $(\partial\G_n^-)$ includes all the regular outgoing 
(incoming) collisions, plus some singular configuration. Excluding the singular points, the collision rule establish an invertible 
and measure preserving transformation between $\partial\G_n^+$ and $\partial\G_n^-$.
Let $\tau_{\pm}(\bz_n) = \inf \{t>0 \mbox{ s.t. } T_{\pm t}^{(n)}\bz_n\in\partial\G_n\}$, i.e. the first forward ($+$) or backwards
($-$) collision time after zero. The connection of $d\s$ with the measure $d\bz_n$ over $\G_n$ is made through the 
map $\bz_n \rightarrow (\bz_n' = T_{-\tau_-(\bz_n)}^{(n)}\bz_n, t' = \tau_-(\bz_n))$, which is one--to--one from 
$\G_n\setminus\partial\G_n$ to the set $\{(\bz_n' ,t')\ \mbox{s.t.}\ \bz_n' \in \partial\G_n^+, t'\in(0,\tau_+(\bz_n'))\}$. 
Namely, we have $d\bz_n = d\tilde\s(\bz_n')dt'$, where $d\tilde\s(\bz_n') = a^2\o\cdot (p_j-p_i)d\s(\bz_n') $ if particle 
$i$ and $j$ are colliding, or $d\tilde\s(\bz_n') = p_i\cdot n(q)d\s(\bz_n') $ if particle $i$ is colliding with the wall.

Remark (1). {\em Any full measure, invariant subset of $\G_n$ intersects $\partial\G_n$ in a set which is full with respect to the 
induced Lebesgue measure.} In fact, if $A\subset\G_n$ is full measure and invariant and $A^c$ is its complement, then 
$0=\int_{A^c}d\bz_n = \int_{A^c \cap \partial\G_n^+} d\tilde\s(\bz'_n)\tau_+(\bz'_n)$. Since the integrand is a.e. strictly positive,
the statement follows.

Remark (2). \label{rem:dyncs}
{\em Any null measure subset of $\G_n$ is avoided by the $n-$particle flow $T^{(n)}_t\bz_n$, for a.a. $(\bz_n,t)\in
\partial\G_n\times\RRR$.} To prove this, we essentially follow \cite{Uchiyama}. Let now $A\subset\G_n$ be a null measure subset.
By the previous remark, points outside $\G_n^*$ are avoided for a.a. $\bz_n\in\partial\G_n$ and all $t$. Hence we may suppose
$A \subset \G_n^*$. For any given $t$, $T^{(n)}_tA$ exists and it is still null measure. But $\int_{T^{(n)}_tA}d\bz_n = 
\int_{\partial\G_n^+\times\RRR^+} d\tilde\s(\bz'_n)dt' \mathbbm{1}_{\{t' < \tau_+(\bz'_n)\}}
\mathbbm{1}_{\{T^{(n)}_{t'}\bz'_n\in T^{(n)}_tA\}}$. This proves the assertion for $t$ restricted to $\RRR^+$. The case $t\in\RRR^-$
is dealed in the same way.

\end{document}